\documentclass[twocolumn,english]{revtex4}
\usepackage[T1]{fontenc}
\usepackage[latin9]{inputenc}
\usepackage{amsmath}
\usepackage{amssymb}
\usepackage{graphicx}
\usepackage{esint}

\makeatletter
\@ifundefined{textcolor}{}
{%
 \definecolor{BLACK}{gray}{0}
 \definecolor{WHITE}{gray}{1}
 \definecolor{RED}{rgb}{1,0,0}
 \definecolor{GREEN}{rgb}{0,1,0}
 \definecolor{BLUE}{rgb}{0,0,1}
 \definecolor{CYAN}{cmyk}{1,0,0,0}
 \definecolor{MAGENTA}{cmyk}{0,1,0,0}
 \definecolor{YELLOW}{cmyk}{0,0,1,0}
}

\usepackage{babel}

\usepackage{babel}

\usepackage{babel}

\usepackage{babel}

\makeatother

\usepackage{babel}
\begin{document}

\title{Magnetic order without tetragonal-symmetry-breaking in iron arsenides:
microscopic mechanism and spin-wave spectrum}

\author{Xiaoyu Wang}

\email{xiaoyu@physics.umn.edu}

\affiliation{School of Physics and Astronomy, University of Minnesota, Minneapolis
55455, USA}

\author{Jian Kang }

\affiliation{School of Physics and Astronomy, University of Minnesota, Minneapolis
55455, USA}

\author{Rafael M. Fernandes}

\affiliation{School of Physics and Astronomy, University of Minnesota, Minneapolis
55455, USA}
\begin{abstract}
Most iron-based superconductors undergo a transition to a magnetically
ordered state characterized by staggered stripes of parallel spins.
With ordering vectors $(\pi,0)$ or $(0,\pi)$, this magnetic state
breaks the high-temperature tetragonal symmetry of the system, which
is manifested by a splitting of the lattice Bragg peaks. Remarkably,
recent experiments in hole-doped iron arsenides reported an ordered
state that displays magnetic Bragg peaks at $(\pi,0)$ and $(0,\pi)$
but remains tetragonal. Despite being inconsistent with a magnetic
stripe configuration, this unusual magnetic phase can be described
in terms of a double-$\mathbf{Q}$ magnetic structure consisting of
an equal-weight superposition of the ordering vectors $(\pi,0)$ and
$(0,\pi)$. Here we show that a non-collinear double-$\mathbf{Q}$
magnetic configuration, dubbed \emph{orthomagnetic}, arises naturally
within an itinerant three-band microscopic model for the iron pnictides.
In particular, we find that strong deviations from perfect nesting
and residual interactions between the electron pockets favor the orthomagnetic
over the stripe magnetic state. Using an effective low-energy model,
we also calculate the spin-wave spectrum of the orthomagnetic state.
In contrast to the stripe state, there are three Goldstone modes,
manifested in all diagonal and one off-diagonal component of the spin-spin
correlation function. The total magnetic structure factor displays
two anisotropic gapless spin-wave branches emerging from both $(\pi,0)$
and $(0,\pi)$ momenta, in contrast to the case of domains of stripe
order, where only one gapless spin-wave branch emerges from each momentum.
We propose that these unique features of the orthomagnetic state can
be used to unambiguously distinguish it from the stripe state via
neutron scattering experiments, and discuss the implications of its
existence to the nature of the magnetism of the iron arsenides. 
\end{abstract}
\maketitle

\section{Introduction}

\begin{figure}
\includegraphics[width=0.8\linewidth]{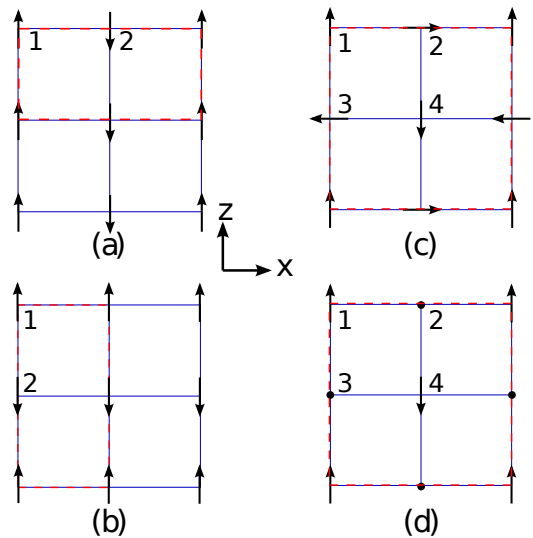}\protect\protect\protect\protect\caption{\label{fig:spin_config}Magnetic ground state configurations for (a)
$\mathbf{Q}_{1}=\left(\pi,0\right)$ stripe order (b) $\mathbf{Q}_{2}=\left(0,\pi\right)$
stripe order (c) double-\textbf{Q} non-collinear (orthomagnetic) order
and (d) double-\textbf{Q} non-uniform magnetic order. The dashed rectangle
denotes the magnetic unit cell in each case. While (a) and (b) are
orthorhombic, (c) and (d) are tetragonal with a unit cell four times
larger than in the paramagnetic phase. }
\end{figure}

Unveiling the nature of the magnetic state of the iron-based materials
\cite{kamihara} is imperative to advance our understanding of their
superconducting state. Indeed, the vast majority of iron arsenide
parent compounds display a stripe magnetically ordered state, characterized
by spins arranged parallel to each other along one in-plane direction
(either the $\mathbf{\hat{x}}$ or the $\hat{\mathbf{y}}$ axis) and
anti-parallel to each other along the other direction, see Fig. \ref{fig:spin_config}a-b
(for reviews, see \cite{reviews}). Its main manifestation is the
presence of magnetic Bragg peaks at the momenta $\mathbf{Q}_{1}=(\pi,0)$
or $\mathbf{Q}_{2}=(0,\pi)$ (in the Fe-square lattice), which correspond
to the ordering vectors of the two possible stripe states. Because
the samples form twin domains, both magnetic peaks are observed in
the same material by neutron scattering experiments \cite{Dai14}.

Since the stripe state breaks the tetragonal $C_{4}$ point-group
symmetry of the system down to the orthorhombic $C_{2}$ symmetry,
a splitting of the lattice Bragg peaks is also observed by x-ray scattering
\cite{Blomberg11}. Remarkably, this orthorhombic distortion is observed
in many systems at a temperature $T_{s}$ above the onset of long-range
magnetic order at $T_{N}$. As a result, an intense debate has been
taking place in the community about the origin of the magnetism in
these materials \cite{Fernandes14,Dagotto12,Eremin14,Johannes09,hu}.
One scenario proposes that the magnetic transition is triggered only
because ferro-orbital order sets in at $T_{s}>T_{N}$, effectively
renormalizing the exchange couplings between neighboring Fe atoms
and enabling magnetic order to be stabilized \cite{lv,w_ku10,Devereaux12}.
A different scenario proposes that the structural transition at $T_{s}$
is a manifestation of an emergent Ising-nematic phase driven by magnetic
fluctuations present near $T_{N}$ \cite{fang,Sachdev,rafael,Dagotto14}.

Recently, new experiments have provided important clues for this hotly
debated topic. In particular, neutron and x-ray scattering measurements
in $\text{Ba}(\text{Fe}_{1-x}\text{Mn}_{x})_{2}\text{As}_{2}$ \cite{kim}
and $(\text{Ba}_{1-x}\text{Na}_{x})\text{Fe}_{2}\text{As}_{2}$ \cite{avci}
reported a regime in which the system displays magnetic Bragg peaks
at $\mathbf{Q}_{1}$ and $\mathbf{Q}_{2}$ but no splitting of the
lattice Bragg peaks. In $(\text{Ba}_{1-x}\text{K}_{x})\text{Fe}_{2}\text{As}_{2}$,
recent thermal expansion measurements observed no orthorhombic distortion
inside the magnetically ordered state near optimal doping \cite{Meiganst_pc}.
Interestingly, in $(\text{Ba}_{1-x}\text{K}_{x})\text{Fe}_{2}\text{As}_{2}$
under pressure, an unidentified ordered state was also observed inside
the magnetic phase \cite{Taillefer12}, which could be connected to
the $C_{4}$-magnetic state found in $(\text{Ba}_{1-x}\text{K}_{x})\text{Fe}_{2}\text{As}_{2}$
at ambient pressure. The only magnetic states compatible with these
reports are double-$\mathbf{Q}$ structures corresponding to a $C_{4}$-preserving
linear combination of the two possible magnetic order parameters \cite{Lorenzana08,Eremin10,Brydon11,xiaoyu}.
Domains of the two different stripe states are incompatible with these
observations, since they would cause a four-fold splitting of the
lattice Bragg peaks \cite{Blomberg11}. More specifically, writing
the spin at position $\mathbf{r}$ as $\mathbf{S}\left(\mathbf{r}\right)=\mathbf{M}_{1}\mathrm{e}^{i\mathbf{Q}_{1}\cdot\mathbf{r}}+\mathbf{M}_{2}\mathrm{e}^{i\mathbf{Q}_{2}\cdot\mathbf{r}}$,
the experimental observations of tetragonal magnetic ground states
in $\text{Ba}(\text{Fe}_{1-x}\text{Mn}_{x})_{2}\text{As}_{2}$, $(\text{Ba}_{1-x}\text{K}_{x})\text{Fe}_{2}\text{As}_{2}$,
and $(\text{Ba}_{1-x}\text{Na}_{x})\text{Fe}_{2}\text{As}_{2}$ imply
a phase with $\left|\mathbf{M}_{1}\right|=\left|\mathbf{M}_{2}\right|$.
If $\mathbf{M}_{1}\perp\mathbf{M}_{2}$, one obtains a non-collinear
phase dubbed orthomagnetic \cite{Lorenzana08} (see Fig. \ref{fig:spin_config}c),
whereas if $\mathbf{M}_{1}\parallel\mathbf{M}_{2}$, a non-uniform
phase emerges where half of the sites are non-magnetic (see Fig. \ref{fig:spin_config}d)
\cite{Eremin10}. In both cases, the system breaks translational symmetry
but not the $C_{4}$ point-group symmetry. The possible existence
of multi-$\mathbf{Q}$ magnetic structures is not a particular feature
of the iron arsenides, as it has also been proposed in a variety of
systems, such as the Kondo system $\mathrm{CeAl_{2}}$ \cite{doubleQ_CeAl2_1,doubleQ_CeAl2_2},
the borocarbide $\mathrm{GdNi_{2}B_{2}C}$ \cite{doubleQ_borocarbides_1,doubleQ_borocarbides_2},
the rocksalt-structure uranium pnictide $\mathrm{USb}$ \cite{tripleQ_USb},
and even $\gamma$-Mn alloys \cite{doubleQ_Mnalloy}.

Taken at face value, the experimental findings of $C_{4}$-preserving
magnetic order in the iron arsenides imply that magnetism can exist
even in the absence of ferro-orbital order, providing indirect evidence
for a magnetic mechanism for the structural transition in the compounds
that display stripe magnetism \cite{avci}. Furthermore, because the
non-collinear and non-uniform states are not present in the ground
state manifold of the local-spin $J_{1}$-$J_{2}$ model \cite{chandra},
their existence favors an itinerant low-energy model for the magnetic
properties of the iron superconductors \cite{Eremin14}. Therefore,
firmly establishing experimentally the existence of these tetragonal
magnetic phases will have a strong impact in our understanding of
these materials. So far, the main evidence in favor of their existence
is the absence of orthorhombic distortion concomitant with the appearance
of magnetic Bragg peaks at $\mathbf{Q}_{1}$ and $\mathbf{Q}_{2}$.
However, x-ray and thermal expansion measurements have an intrinsic
resolution limitation that could render the detection of very small
lattice distortions difficult \cite{inosov,Osborn14}. Furthermore,
at least in the case of $\text{Ba}(\text{Fe}_{1-x}\text{Mn}_{x})_{2}\text{As}_{2}$,
it has been proposed that disorder effects related to the Mn doping
could account for some of the puzzling experimental observations \cite{gastiasoro,inosov}.
Thus, it is desirable to search for other unambiguous experimental
signatures of these $C_{4}$-magnetic phases. While previous works
have focused on the signatures of the non-uniform double-$\mathbf{Q}$
phase \cite{xiaoyu}, the properties of the orthomagnetic state remain
largely unexplored.

In this paper, we show explicitly from a microscopic three-band model
that strong deviations from particle-hole symmetry (perfect nesting)
favor a tetragonal magnetic state over the stripe state. Interestingly,
the orthomagnetic state is selected by a residual electronic interaction
that does not participate explicitly in the formation of the magnetic
state \cite{Eremin10}. These theoretical results are complementary
to those reported in Ref. \cite{avci}, which found that deep inside
the stripe ordered state a second instability towards a tetragonal
magnetic state emerges.

Using an effective low-energy model for the orthomagnetic state, we
also calculate its spin-wave spectrum. In contrast to the stripe state,
which displays a doubly-degenerate Goldstone mode, we find three Goldstone
modes that give rise to two distinct spin-wave branches emerging from
both $\mathbf{Q}_{1}$ and $\mathbf{Q}_{2}$. Furthermore, the three
diagonal components of the spin-spin correlation function $\mathcal{S}_{xx}$,
$\mathcal{S}_{yy}$, $\mathcal{S}_{zz}$, and the off-diagonal $\mathcal{S}_{xz}$
term, display spin-wave modes at low energies, implying that in all
directions the excitations behave as gapless transverse-like modes,
as expected for a non-collinear configuration that breaks all spin-space
rotational symmetries. We argue that these signatures of the low-energy
spin spectrum of the orthomagnetic state can be unambiguously distinguished
from those arising from domains of stripe magnetic states in both
unpolarized and polarized neutron scattering measurements. Thus, our
studies provide concrete criteria to establish the existence of the
orthomagnetic state beyond the indirect evidence of an absent orthorhombic
distortion.

Our paper is organized as follows: in Section II, we review the microscopic
model for the magnetic instability proposed in Refs. \cite{rafael,Eremin10}
and extend the analysis by including the effects of the residual electronic
interactions and by computing the Ginzburg-Landau coefficients in
the regime far from perfect nesting. In Section III we present an
effective low-energy model for the orthomagnetic state whose spin-wave
dispersions can be appropriately treated within the Holstein-Primakoff
formalism. We calculate both the spin-wave modes and the components
of the spin-spin correlation function, contrasting to the stripe magnetic
case. Section IV is devoted to the concluding remarks and to the applicability
of our results to other materials that may display orthomagnetic order,
such as the heavy-fermion related compound GdRhIn$_{5}$ \cite{granado}.

\section{Microscopic mechanism for the orthomagnetic order}

We start with the itinerant three-band model that was previously used
to explain the magnetic properties of the iron arsenides near perfect
nesting \cite{rafael,Eremin10}. The non-interacting part consists
of:

\begin{equation}
H_{0}=\sum_{\mathbf{k}a\alpha}\varepsilon_{a,\mathbf{k}}c_{a,\mathbf{k}\alpha}^{\dagger}c_{a,\mathbf{k}\alpha}\label{H0}
\end{equation}
where $\alpha$ is the spin index, $\mathbf{k}$ is the momentum,
and $a=h$, $e_{1}$, $e_{2}$ refer to, respectively, the central
hole pocket and the electron pockets centered at $\mathbf{Q}_{1}=\left(\pi,0\right)$
and $\mathbf{Q}_{2}=\left(0,\pi\right)$. The band dispersions near
the Fermi level can be conveniently parametrized as:

\begin{align}
\varepsilon_{h,\mathbf{k}} & =-\varepsilon_{\mathbf{k}}\nonumber \\
\varepsilon_{e_{1},\mathbf{k}+\mathbf{Q}_{1}} & =\varepsilon_{\mathbf{k}}-\left(\delta_{\mu}+\delta_{m}\cos2\theta\right)\nonumber \\
\varepsilon_{e_{2},\mathbf{k}+\mathbf{Q}_{2}} & =\varepsilon_{\mathbf{k}}-\left(\delta_{\mu}-\delta_{m}\cos2\theta\right)\label{aux_H0}
\end{align}
where $\varepsilon_{\mathbf{k}}=\frac{k^{2}}{2m}-\varepsilon_{0}$
is a parabolic-like band dispersion, $\theta$ is the angle around
the electron pocket, $\delta_{m}$ describes the ellipticity of the
electron pockets, and $\delta_{\mu}$ is proportional to changes in
the carrier concentration (doping). Note that for $\delta_{m}=\delta_{\mu}=0$,
the system has perfect particle-hole symmetry and the hole and electron
pockets are perfectly nested.

Following Ref. \cite{chubukov,maiti}, there are eight types of purely
electronic interactions connecting the three Fermi pockets, corresponding
to density-density ($U_{1}$, $U_{4}$, $U_{5}$, $U_{6}$), spin-exchange
($U_{2}$, $U_{7}$) and pair-hopping ($U_{3}$, $U_{8}$) interactions
respectively. The interaction Hamiltonian is: 
\begin{equation}
\begin{split}H_{\text{int}} & =U_{1}\sum c_{h\alpha}^{\dagger}c_{e_{i}\beta}^{\dagger}c_{e_{i}\beta}c_{h\alpha}+U_{2}\sum c_{h\alpha}^{\dagger}c_{e_{i}\beta}^{\dagger}c_{h\beta}c_{e_{i}\alpha}\\
 & +\frac{U_{3}}{2}\sum(c_{h\alpha}^{\dagger}c_{h\beta}^{\dagger}c_{e_{i}\beta}c_{e_{i}\alpha}+\mathrm{h.c.})\\
 & +\frac{U_{4}}{2}\sum c_{e_{i}\alpha}^{\dagger}c_{e_{i}\beta}^{\dagger}c_{e_{i}\beta}c_{e_{i}\alpha}+\frac{U_{5}}{2}\sum c_{h\alpha}^{\dagger}c_{h\beta}^{\dagger}c_{h\beta}c_{h\alpha}\\
 & +U_{6}\sum c_{e_{1}\alpha}^{\dagger}c_{e_{2}\beta}^{\dagger}c_{e_{2}\beta}c_{e_{1}\alpha}+U_{7}\sum c_{e_{1}\alpha}^{\dagger}c_{e_{2}\beta}^{\dagger}c_{e_{1}\beta}c_{e_{2}\alpha}\\
 & +\frac{U_{8}}{2}\sum(c_{e_{1}\alpha}^{\dagger}c_{e_{1}\beta}^{\dagger}c_{e_{2}\beta}c_{e_{2}\alpha}+\mathrm{h.c.})
\end{split}
\label{eq:int_hamltonian}
\end{equation}

For simplicity of notation, the momentum indices are all suppressed
with the implicit constraint of momentum conservation. To study the
instability towards magnetic order, we project all the interactions
in the spin-density wave (SDW) channel -- which is the leading one
according to RG and fRG calculations \cite{maiti,Thomale09}. The
only interactions that contribute directly to the SDW instability
are $U_{1}$ and $U_{3}$. The partition function, restricted to this
channel only, can then be written in the functional field form: 
\begin{equation}
\mathcal{Z}=\int\mathcal{D}c^{\dagger}\mathcal{D}c\ \exp(-S)\label{aux_Z}
\end{equation}
with the action : 
\begin{equation}
S=\int_{0}^{\beta}\mathrm{d}\tau\ \sum_{i\mathbf{k}\sigma}c_{i\mathbf{k}\sigma}^{\dagger}\partial_{\tau}c_{i\mathbf{k}\sigma}+H_{0}+H_{\mathrm{SDW}}\label{aux_S}
\end{equation}
and the SDW-decoupled interaction: 
\begin{equation}
H_{\text{SDW}}=-I\sum_{i\mathbf{k}\mathbf{k}^{\prime}\mathbf{q}}\left(c_{h\mathbf{k}\alpha}^{\dagger}\boldsymbol{\sigma}_{\alpha\beta}c_{e_{i}\mathbf{k}+\mathbf{q}\beta}\right)\cdot\left(c_{e_{i}\mathbf{k}^{\prime}\gamma}^{\dagger}\boldsymbol{\sigma}_{\gamma\delta}c_{h\mathbf{k}^{\prime}-\mathbf{q}\delta}\right)\label{HSDW}
\end{equation}
where $I=U_{1}+U_{3}$. We now introduce the Hubbard-Stratonovich
fields $\mathbf{M}_{i}$, whose mean value is proportional to the
staggered magnetization with ordering vector $\mathbf{Q}_{i}$, i.e.
$\left\langle \mathbf{M}_{i}\right\rangle =I\sum_{\mathbf{k}}\left\langle c_{h\mathbf{k}\alpha}^{\dagger}\boldsymbol{\sigma}_{\alpha\beta}c_{e_{i}\mathbf{k}+\mathbf{q}\beta}\right\rangle $,
via $\exp\left(-H_{\text{SDW}}\right)\propto\int\mathcal{D}\mathbf{M}_{i}\exp\left(-S_{\mathrm{SDW}}\left[\mathbf{M}_{i}\right]\right)$
with:

\begin{align}
S_{\mathrm{SDW}}\left[\mathbf{M}_{i}\right] & =\sum_{i\mathbf{q}}\frac{\mathbf{M}_{i\mathbf{q}}\cdot\mathbf{M}_{i-\mathbf{q}}}{I}\label{aux}\\
 & -\sum_{i\mathbf{q}\mathbf{k}}\left(\mathbf{M}_{i\mathbf{q}}\cdot c_{e_{i}\mathbf{k}\gamma}^{\dagger}\boldsymbol{\sigma}_{\gamma\delta}c_{h\mathbf{k}-\mathbf{q}\delta}+\mathrm{h.c.}\right)\nonumber 
\end{align}

Following Ref. \cite{rafael}, we then integrate out the electronic
degrees of freedom, obtaining an effective action for the magnetic
degrees of freedom:

\begin{align}
\mathcal{Z} & =\int\mathcal{D}c^{\dagger}\mathcal{D}c\,\mathcal{D}\mathbf{M}_{i}\ \exp\left(-S\left[c^{\dagger},\mathbf{M}_{i}\right]\right)\nonumber \\
 & =\int\mathcal{D}\mathbf{M}_{i}\ \exp\left(-S_{\mathrm{eff}}\left[\mathbf{M}_{i}\right]\right)\label{Z_eff}
\end{align}

For a finite-temperature magnetic transition, $S_{\mathrm{eff}}\left[\mathbf{M}_{i}\right]=F\left[\mathbf{M}_{i}\right]/T$,
where $F\left[\mathbf{M}_{i}\right]$ is the free energy. Near the
magnetic transition, we can expand the action in powers of the magnetic
order parameters, deriving the Ginzburg-Landau expansion:

\begin{equation}
\begin{split}S_{\mathrm{eff}} & \left[\mathbf{M}_{i}\right]=\frac{a}{2}(\mathbf{M}_{1}^{2}+\mathbf{M}_{2}^{2})+\frac{u}{4}(\mathbf{M}_{1}^{2}+\mathbf{M}_{2}^{2})^{2}\\
 & -\frac{g}{4}(\mathbf{M}_{1}^{2}-\mathbf{M}_{2}^{2})^{2}+w(\mathbf{M}_{1}\cdot\mathbf{M}_{2})^{2}
\end{split}
\label{free_energy}
\end{equation}

In the vicinity of the magnetic transition, $a\approx N_{f}(T-T_{N})$,
where $N_{f}$ is the density of states at the Fermi surface. The
coefficients $u$, $g$, and $w$ are given by \cite{rafael}:

\begin{align}
u & =A+B\nonumber \\
g & =B-A\nonumber \\
w & =0\label{u_g_w}
\end{align}
with:

\begin{align}
A & =\int_{k}G_{h,\mathbf{k}}^{2}G_{e_{1},\mathbf{k}+\mathbf{Q}_{1}}^{2}\nonumber \\
 & =\int_{k}\left(\frac{1}{i\omega_{n}+\varepsilon_{\mathbf{k}}}\right)^{2}\left(\frac{1}{i\omega_{n}-\varepsilon_{\mathbf{k}}+\delta_{\mu}+\delta_{m}\cos2\theta}\right)^{2}\quad,\nonumber \\
B & =\int_{k}G_{h,\mathbf{k}}^{2}G_{e_{1},\mathbf{k+Q_{1}}}G_{e_{2},\mathbf{k}+\mathbf{Q}_{2}}\nonumber \\
 & =\int_{k}\left(\frac{1}{i\omega_{n}+\varepsilon_{\mathbf{k}}}\right)^{2}\frac{1}{i\omega_{n}-\varepsilon_{\mathbf{k}}+\delta_{\mu}+\delta_{m}\cos2\theta}\nonumber \\
 & \ \times\frac{1}{i\omega_{n}-\varepsilon_{\mathbf{k}}+\delta_{\mu}-\delta_{m}\cos2\theta}\quad,\label{aux_A_B}
\end{align}

Here, $G_{a,\mathbf{k}}$ is the non-interacting fermionic Green's
function for pocket $a$, $G_{a,\mathbf{k}}^{-1}=i\omega_{n}-\varepsilon_{a,\mathbf{k}}$,
and $\int_{k}\rightarrow T\sum_{n}\int\frac{d\mathbf{k}}{\left(2\pi\right)^{d}}$,
with Matsubara frequency $\omega_{n}=\left(2n+1\right)\pi T$. After
integrating out the momentum, we obtain:

\begin{align}
A & =N_{f}T\pi\sum_{n=0}^{\infty}\mathrm{Im}\int_{\theta}\frac{1}{\left(i\omega_{n}+\frac{\delta_{\mu}}{2}+\frac{\delta_{m}}{2}\,\cos2\theta\right)^{3}}\nonumber \\
B & =N_{f}T\pi\sum_{n=0}^{\infty}\mathrm{Im}\int_{\theta}\frac{i\omega_{n}+\frac{\delta_{\mu}}{2}}{\left(\left(i\omega_{n}+\frac{\delta_{\mu}}{2}\right)^{2}-\left(\frac{\delta_{m}}{2}\,\cos2\theta\right)^{2}\right)^{2}}\label{eq_A_B}
\end{align}
with $\int_{\theta}=\int_{0}^{2\pi}\frac{d\theta}{2\pi}$. A straightforward
minimization of Eq. (\ref{free_energy}) reveals that the stripe magnetic
state is the global free energy minimum for $g>\max\left(0,-w\right)$,
whereas a tetragonal magnetic state is the lowest energy ground state
for $g<\max\left(0,-w\right)$. Since in the model above $w=0$, the
sign of $g$ determines uniquely the symmetry of the magnetic ground
state. Note that the free energy functional is bounded for $u>\max\left(g,0,-w\right)$.

\begin{figure}[htb]
\begin{centering}
\includegraphics[width=0.7\linewidth]{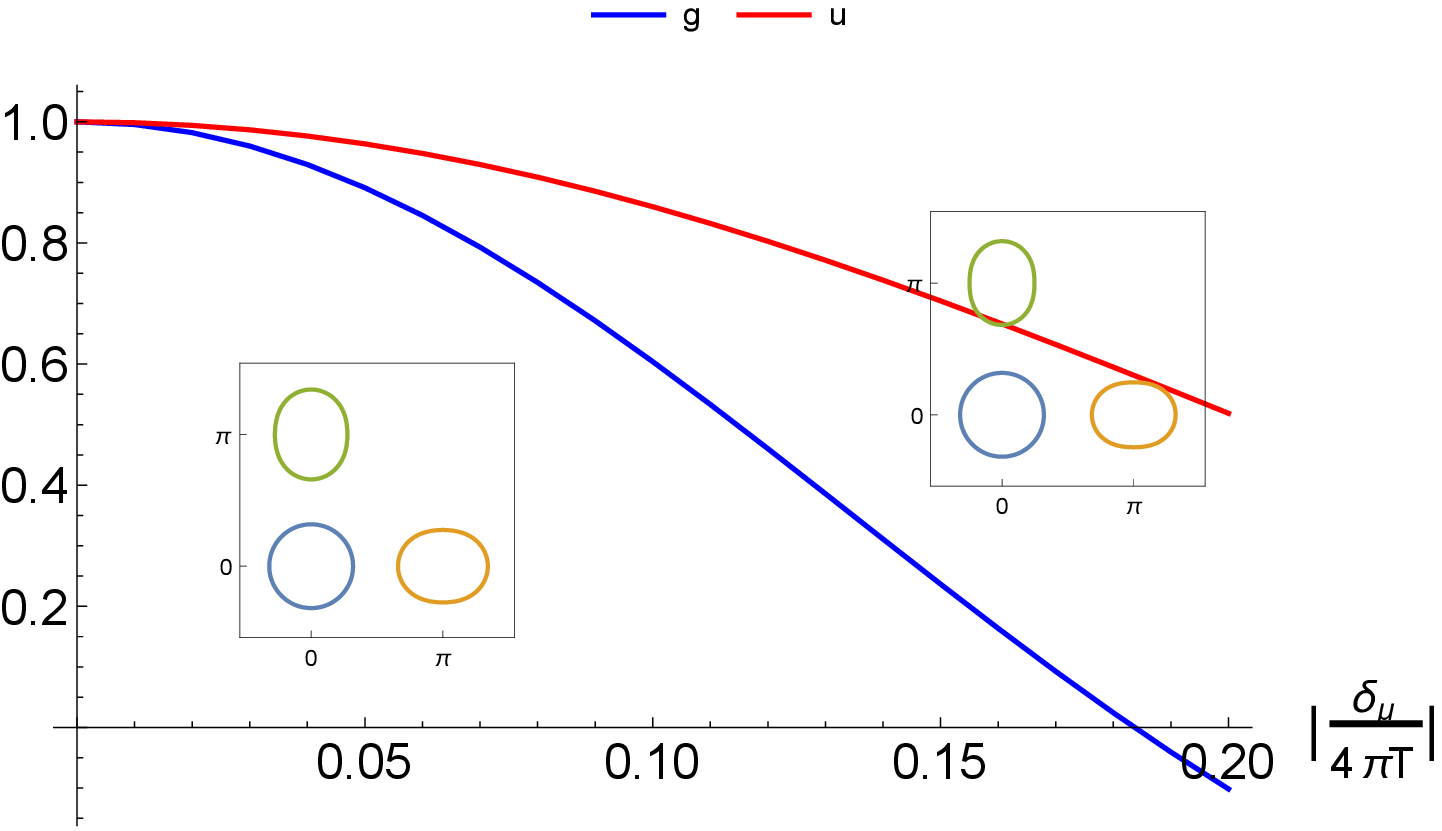} 
\par\end{centering}

\medskip{}

\begin{centering}
\includegraphics[width=0.7\linewidth]{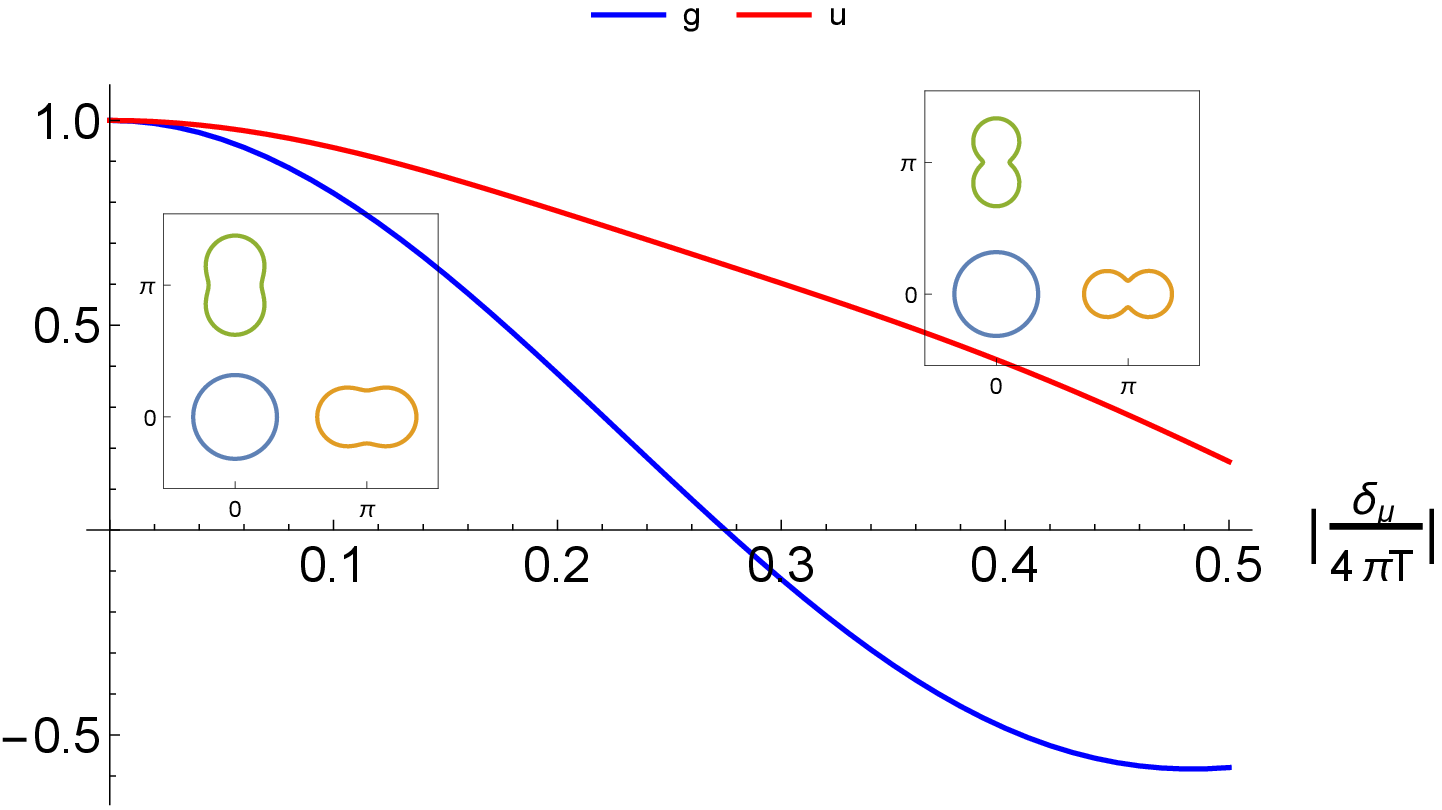} 
\par\end{centering}

\protect\protect\caption{Quartic Ginzburg-Landau coefficients $u$ (red) and $g$ (blue) of
the free energy (\ref{free_energy}) as function of $|\delta_{\mu}|/(4\pi T)$
for $\delta_{m}/(4\pi T)=0.2$ (top) and $\delta_{m}/(4\pi T)=0.5$
(bottom). The insets show the shape of the Fermi pockets for $\frac{\delta_{\mu}}{4\pi T}=-0.05\ \text{and}\ -0.2$
in the top panel, $\frac{\delta_{\mu}}{4\pi T}=-0.1\ \text{and}\ -0.4$
in the bottom panel. Note that $u$ and $g$ are normalized by their
values at $\delta_{\mu}=0$. For $g>0$, the magnetic ground state
is the stripe state, which lowers the tetragonal symmetry to orthorhombic.
For $g<0$, tetragonal symmetry is preserved, and either a non-collinear
or a non-uniform double-$\mathbf{Q}$ magnetic state arises. Notice
that the free energy remains bounded as long as $u>0$. }

\label{Fig::GNoSC} 
\end{figure}

The previous analysis of the Ginzburg-Landau coefficients (\ref{eq_A_B})
in Ref. \cite{rafael} focused on the regime near perfect nesting,
where $\delta_{\mu},\delta_{m}\ll T$. In this case, $g\propto\delta_{m}^{2}>0$
and the ground state is the stripe magnetic one. Here, we extend the
analysis beyond small deviations from perfect nesting by numerically
computing Eqs. (\ref{eq_A_B}) for arbitrary $\delta_{\mu}/T$, $\delta_{m}/T$
(the constraint $\delta_{m}\geq\delta_{\mu}$ is imposed to ensure
that hot spots are present, as seen experimentally). To mimic the
phase diagrams of $\text{Ba}(\text{Fe}_{1-x}\text{Mn}_{x})_{2}\text{As}_{2}$,
$(\text{Ba}_{1-x}\text{K}_{x})\text{Fe}_{2}\text{As}_{2}$, and $(\text{Ba}_{1-x}\text{Na}_{x})\text{Fe}_{2}\text{As}_{2}$,
we change the parameter $\delta_{\mu}$ (proportional to the carrier
concentration) for a fixed value of the ellipticity $\delta_{m}$.
Note that Eq. (\ref{eq_A_B}) implies that the behavior of $u$ and
$g$ depend only on $|\delta_{\mu}|$. Fig \ref{Fig::GNoSC} shows
the results for $\delta_{m}/T=0.8\pi$ and $\delta_{m}/T=2\pi$. For
small values of $|\delta_{\mu}|$, both $u$ and $g$ are positive,
and the stripe magnetic state is favored. However, as $|\delta_{\mu}|$
becomes larger, regardless of the value of $\delta_{m}$, $g$ becomes
negative, indicating that the magnetic ground state becomes a double-$\mathbf{Q}$
tetragonal phase. The evolution of the Fermi surfaces as $|\delta_{\mu}|$
increases, for the case of hole-doping, is shown in the insets. Since
$u$ remains positive when $g$ first changes sign, the free energy
remains bounded, i.e. the mean-field transition is second-order. These
results are in qualitative agreement with the phase diagrams of $\text{Ba}(\text{Fe}_{1-x}\text{Mn}_{x})_{2}\text{As}_{2}$,
$(\text{Ba}_{1-x}\text{K}_{x})\text{Fe}_{2}\text{As}_{2}$, and $(\text{Ba}_{1-x}\text{Na}_{x})\text{Fe}_{2}\text{As}_{2}$,
which display the tetragonal magnetic phase only for sufficiently
strong doping concentration. We note that a tetragonal magnetic state
was also reported in other itinerant approaches for the iron pnictides
\cite{Eremin10,Lorenzana08,Brydon11}, as well as in a strong-coupling
two-orbital ladder model \cite{Berg10}.

\begin{figure}
\includegraphics[width=0.9\linewidth]{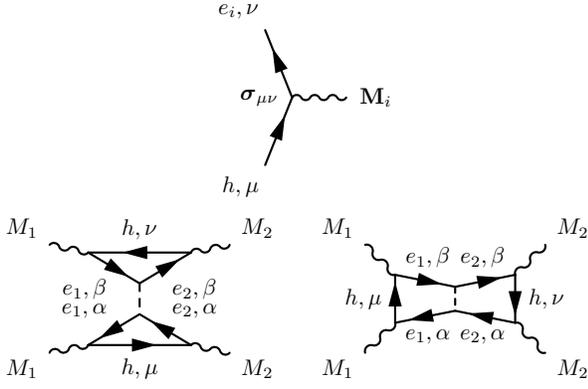} \protect\protect\protect\protect\caption{Top panel: The vertex that couples the magnetic order parameter to
the low-energy fermionic states, which has a $\mathbf{M}_{i}\cdot\boldsymbol{\sigma}$
structure. Solid lines refer to the non-interacting electronic Green's
functions. Bottom panel: the Feynman diagrams containing the leading-order
corrections to the free energy arising from the residual $U_{7}$
interaction. \label{fig_ee_interaction} }
\end{figure}

Because $w=0$ in our model, when $g$ becomes negative the system
does not distinguish between the two possible tetragonal magnetic
states, namely, the orthomagnetic one ($w>0$, $\mathbf{M}_{1}\perp\mathbf{M}_{2}$)
and the non-uniform one ($w<0$, $\mathbf{M}_{1}\parallel\mathbf{M}_{2}$).
Although the terms arising purely from the band structure do not contribute
to the $w$ coefficient, the residual interactions in Eq. (\ref{eq:int_hamltonian})
that do not participate in the SDW instability (namely $U_{2}$, $U_{4}$,
$U_{5}$, $U_{6}$, $U_{7}$, and $U_{8}$) give rise to such a term,
as pointed out in Ref. \cite{Eremin10}. Computing the contributions
of the residual interactions to the action, we obtain:

\begin{align}
\tilde{S}\left[\mathbf{M}_{i}\right] & =\frac{1}{4}\left(-\tilde{u}_{2}+\tilde{u}_{4}+\tilde{u}_{5}+\tilde{u}_{6}-\tilde{u}_{7}+\tilde{u}_{8}\right)\left(\mathbf{M}_{1}^{2}+\mathbf{M}_{2}^{2}\right)^{2}\nonumber \\
 & -\frac{1}{4}\left(-\tilde{g}_{4}+\tilde{g}_{6}-\tilde{g}_{7}+\tilde{g}_{8}\right)\left(\mathbf{M}_{1}^{2}-\mathbf{M}_{2}^{2}\right)^{2}\nonumber \\
 & +\tilde{w}_{7}\left(\mathbf{M}_{1}\cdot\mathbf{M}_{2}\right)^{2}\label{F_corrections}
\end{align}

The Ginzburg-Landau coefficients can be obtained in a straightforward
way using diagrammatics. In Fig. \ref{fig_ee_interaction} we show
the two diagrams arising from the $U_{7}$ interaction, which contribute
to $\tilde{u}_{7}$, $\tilde{g}_{7}$, and $\tilde{w}_{7}$. Additional
computational details are discussed in Appendix A. The coefficients
of the quartic symmetric term are given by:

\begin{align}
\tilde{u}_{2} & =8U_{2}\left(\int_{k}G_{h,\mathbf{k}}G_{e_{1},\mathbf{k}+\mathbf{Q}_{1}}^{2}\right)\left(\int_{k}G_{h,\mathbf{k}}^{2}G_{e_{1},\mathbf{k}+\mathbf{Q}_{1}}\right)\nonumber \\
\tilde{u}_{4} & =2U_{4}\left(\int_{k}G_{h,\mathbf{k}}G_{e_{1},\mathbf{k}+\mathbf{Q}_{1}}^{2}\right)^{2}\nonumber \\
\tilde{u}_{5} & =4U_{5}\left(\int_{k}G_{h,\mathbf{k}}^{2}G_{e_{1},\mathbf{k}+\mathbf{Q}_{1}}\right)^{2}\nonumber \\
\tilde{u}_{6} & =2U_{6}\left[2\left(\int_{k}G_{h,\mathbf{k}}G_{e_{1},\mathbf{k}+\mathbf{Q}_{1}}^{2}\right)^{2}\right.\nonumber \\
 & \left.-\left(\int_{k}G_{h,\mathbf{k}}G_{e_{1},\mathbf{k}+\mathbf{Q}_{1}}G_{e_{2},\mathbf{k}+\mathbf{Q}_{2}}\right)^{2}\right]\nonumber \\
\tilde{u}_{7} & =2U_{7}\left(\int_{k}G_{h,\mathbf{k}}G_{e_{1},\mathbf{k}+\mathbf{Q}_{1}}^{2}\right)^{2}\nonumber \\
\tilde{u}_{8} & =2U_{8}\left(\int_{k}G_{h,\mathbf{k}}G_{e_{1},\mathbf{k}+\mathbf{Q}_{1}}G_{e_{2},\mathbf{k}+\mathbf{Q}_{2}}\right)^{2}\label{GL_corrections_u}
\end{align}
whereas the coefficients of the quartic anti-symmetric term read:

\begin{align}
\tilde{g}_{4} & =2U_{4}\left(\int_{k}G_{h,\mathbf{k}}G_{e_{1},\mathbf{k}+\mathbf{Q}_{1}}^{2}\right)^{2}\nonumber \\
\tilde{g}_{6} & =2U_{6}\left[2\left(\int_{k}G_{h,\mathbf{k}}G_{e_{1},\mathbf{k}+\mathbf{Q}_{1}}^{2}\right)^{2}\right.\nonumber \\
 & \left.-\left(\int_{k}G_{h,\mathbf{k}}G_{e_{1},\mathbf{k}+\mathbf{Q}_{1}}G_{e_{2},\mathbf{k}+\mathbf{Q}_{2}}\right)^{2}\right]\nonumber \\
\tilde{g}_{7} & =2U_{7}\left(\int_{k}G_{h,\mathbf{k}}G_{e_{1},\mathbf{k}+\mathbf{Q}_{1}}^{2}\right)^{2}\nonumber \\
\tilde{g}_{8} & =2U_{8}\left(\int_{k}G_{h,\mathbf{k}}G_{e_{1},\mathbf{k}+\mathbf{Q}_{1}}G_{e_{2},\mathbf{k}+\mathbf{Q}_{2}}\right)^{2}\label{GL_corrections_g}
\end{align}
and the coefficient of the quartic scalar-product term yields:

\begin{equation}
\tilde{w}_{7}=4U_{7}\left(\int_{k}G_{h,\mathbf{k}}G_{e_{1},\mathbf{k}+\mathbf{Q}_{1}}G_{e_{2},\mathbf{k}+\mathbf{Q}_{2}}\right)^{2}\label{GL_corrections_w}
\end{equation}

At perfect nesting, an overall factor proportional to the Green's
functions product $\left(\int_{k}G_{h,\mathbf{k}}^{2}G_{e_{1},\mathbf{k}+\mathbf{Q}_{1}}\right)^{2}$
appears in all terms. In this limit, our results become identical
to those found in Ref. \cite{Eremin10}, which computed the corrections
to the magnetic ground state energy in the ordered state using a sequence
of Bogoliubov transformations. In the paramagnetic state, however,
which is our case of interest, this Green's functions product vanishes
-- as also pointed out in Ref. \cite{Eremin10}. The corrections due
to the residual interactions naturally become non-zero -- and in fact
positive -- once one considers small deviations from perfect nesting.
For $\tilde{w}_{7}$, we find:

\begin{equation}
\tilde{w}_{7}\approx4U_{7}\left(\frac{7\zeta(3)N_{f}\delta_{\mu}}{8\pi^{2}T^{2}}\right)^{2}\label{expansion_w7}
\end{equation}

Because the band dispersions do not contribute to the term $\left(\mathbf{M}_{1}\cdot\mathbf{M}_{2}\right)^{2}$,
the fact that $\tilde{w}_{7}>0$ is very important, as it implies
that the residual interaction $U_{7}$ selects the orthomagnetic state
over the non-uniform state (assuming that $U_{7}>0$, as one would
expect). Therefore, when $g_{\mathrm{eff}}=g+\tilde{g}$ changes sign,
the system tends to form the non-collinear tetragonal magnetic state
shown in Fig. \ref{fig:spin_config}c. Dimensional analysis of the
relevant Feynman diagrams reveals that $\tilde{g}/g\propto U_{i}N_{f}$,
where $U_{i}$ is the appropriate combination of residual interactions.
Therefore, in our weak-coupling approach, because $U_{i}N_{f}\ll1$,
it follows that $\tilde{g}\ll g$ -- unless $g$ itself is close to
zero. As a result, although the contribution from $\tilde{g}$ may
change slightly the value of $\delta_{\mu}$ for which $g_{\mathrm{eff}}=g+\tilde{g}$
vanishes, it cannot preclude the sign-changing found in Fig. \ref{Fig::GNoSC}
from taking place. The case of $w_{\mathrm{eff}}=w+\tilde{w}$ is
fundamentally different, since $w=0$, making $\tilde{w}$ the leading
non-vanishing term.

In the next section, we will discuss the magnetic spectrum of such
a state. Before proceeding, we emphasize that, as pointed out by two
of us in a previous communication \cite{xiaoyu}, other mechanisms
may favor a different sign for the $w$ coefficient -- such as the
coupling to soft Neel-like magnetic fluctuations -- which could stabilize
the non-uniform tetragonal magnetic state shown in Fig. \ref{fig:spin_config}d.
We also note that the Ginzburg-Landau expansion (\ref{free_energy})
is very general for two SDW order parameters that preserve spin-rotational
and tetragonal symmetries. To obtain its coefficients, besides the
Hertz-Millis approach employed here, one can also fit the free energy
directly to first-principle band structures. This was done in Ref.
\cite{giovannetti}, which also found the orthomagnetic state to be
a ground state for certain parameter ranges.

\section{Spin-wave spectrum}

Having established the conditions under which the orthomagnetic state
becomes the ground state of the system, we now discuss its experimental
manifestations. The most evident one is the lack of tetragonal symmetry
breaking, since the orthomagnetic order has an equal weight of the
order parameters $\mathbf{M}_{1}$ and $\mathbf{M}_{2}$ associated
with the ordering vectors $\mathbf{Q}_{1}=\left(\pi,0\right)$ and
$\mathbf{Q}_{2}=\left(0,\pi\right)$, respectively. The preservation
of $C_{4}$ symmetry can in principle be detected by x-ray or neutron
scattering via the absence of splitting of the lattice Bragg peaks
across the magnetic transition. However, given the resolution limitations
of scattering measurements, it is desirable to consider other properties
that identify unambiguously the orthomagnetic state.

In this section, we study in details the spin-wave spectrum of the
orthomagnetic phase, comparing it to the stripe phase. As we are interested
in the low-energy behavior, there are two alternative approaches to
compute the spin-wave spectrum: the first is by evaluating self-consistently
the poles of the spin-spin correlation function deep inside the magnetically
ordered state within the itinerant approach described in the previous
section \cite{spin_waves_itinerant_MSDW,knolle01,knolle02}. The second
alternative is to build a phenomenological localized-spin model that
gives the same ground states as the itinerant model, and then use
Holstein-Primakoff (HP) bosons to compute the spin-wave dispersion\cite{zhao,harriger}.
Given the simplicity of the latter, we here consider a Heisenberg
model on a two-dimensional square lattice, with nearest-neighbor and
next-nearest neighbor interactions: 
\begin{equation}
H=J_{1}\sum_{\langle i,j\rangle}\mathbf{S}_{i}\cdot\mathbf{S}_{j}+J_{2}\sum_{\langle\langle i,j\rangle\rangle}\mathbf{S}_{i}\cdot\mathbf{\mathbf{S}}_{j}-\frac{K}{S^{2}}\sum_{\langle i,j\rangle}(\mathbf{S}_{i}\cdot\mathbf{S}_{j})^{2}\label{eq:heisenberg_model}
\end{equation}
where $\langle...\rangle$ and $\langle\langle...\rangle\rangle$
denote nearest-neighbors and next-nearest neighbors, and $J_{1}>0$,
$J_{2}>J_{1}/2$ are the respective antiferromagnetic exchange interactions.
The biquadratic term $K$ selects between the stripe phase ($K>0$)
and the orthomagnetic phase ($K<0$) in the classical regime. We emphasize
that this is a phenomenological model constructed to describe the
ground states obtained in Section II. Indeed, if it was the classical
$J_{1}$-$J_{2}$ model, $K$ would be restricted to small positive
values only \cite{chandra}. Instead, here $K$ should be understood
as a phenomenological parameter, analogous to the parameter $g$ calculated
in Eq. (\ref{free_energy}). In fact, a Ginzburg-Landau expansion
of this toy Heisenberg model would result in a free energy equivalent
to that of Eq. (\ref{free_energy}), evidencing the fact that both
models share the same low-energy properties \cite{Batista11}. Therefore,
the use of this localized-spin model should be understood simply as
a tool to evaluate the spin-wave spectrum, and not an implication
that local moments are necessarily present in the system. Incidentally,
we note that other Heisenberg models with ring exchange interactions
can also display orthomagnetic order \cite{Chubukov92}.

We emphasize that a strict two-dimensional model does not have long-range
Heisenberg magnetic order, according to Mermin-Wagner theorem. As
a result, we assume here that the system is formed by weakly-coupled
layers. Such a small inter-layer coupling can nevertheless be neglected
in what regards the main properties of the spin-wave dispersions.
To obtain the spin-wave spectrum of the Hamiltonian (\ref{eq:heisenberg_model}),
we follow Refs. \cite{Carlson04,haraldsen} and introduce locally
Holstein-Primakoff (HP) bosons for each of the $r$ spins in a single
magnetic unit cell: 
\begin{align}
\mathbf{S}_{z}^{(r)} & =S-a^{(r)\dagger}a^{(r)}\nonumber \\
\mathbf{S}_{+}^{(r)} & =\sqrt{2S-a^{(r)\dagger}a^{(r)}}\ a^{(r)}\label{eq:hp_trans}\\
\mathbf{S}_{-}^{(r)} & =a^{(r)\dagger}\sqrt{2S-a^{(r)\dagger}a^{(r)}}\nonumber 
\end{align}

Note that the spin coordinate system is defined locally, such that
the local spin is always parallel to the local $+\hat{\mathbf{z}}$
axis. For convenience, the two-dimensional lattice plane $xy$ is
chosen to be the $xz$ spin-plane, as shown in Fig. \ref{fig:spin_config}.
Since different types of spins within a magnetic unit cell have their
own degree of freedom, the number of HP bosons (labeled by $r$) is
equal to the number of spins within a magnetic unit cell. Thus, the
stripe state has $r=1,2$ whereas the orthomagnetic state has $r=1,2,3,4$,
as labeled in Fig. \ref{fig:spin_config}. The Fourier transform of
the HP bosons is defined as 
\begin{equation}
a_{\mathbf{k}}^{(r)}=\sum_{i\in r}a_{i}^{(r)}e^{-i\mathbf{k}\cdot\mathbf{x}_{i}^{(r)}}\label{eq_a}
\end{equation}
where $i$ labels different magnetic unit cells, and $\mathbf{x}_{i}^{(r)}$
is the position of the $r$-th spin in the $i$-th magnetic unit cell.
For convenience, we define: 
\begin{equation}
\Psi_{\mathbf{k}}^{\dagger}\equiv\left(a_{\mathbf{k}}^{(1)\dagger},a_{-\mathbf{k}}^{(1)},a_{\mathbf{k}}^{(2)\dagger},a_{-\mathbf{k}}^{(2)},...\right)\label{eq:psi}
\end{equation}

Because we are interested in the classical limit, we perform a large
$S$ expansion and keep only terms that are quadratic in the bosonic
operators. In this case, the Heisenberg Hamiltonian can be re-expressed
as: 
\begin{equation}
H=H_{0}+\frac{1}{2}\sum_{\mathbf{k}}\Psi_{\mathbf{k}}^{\dagger}\hat{\mathcal{H}}_{\mathbf{k}}\Psi_{\mathbf{k}}\label{eq:qp_repres}
\end{equation}
where $H_{0}$ is the classical ground state energy for a given spin
configuration. The spin-wave modes can be obtained by a generalized
Bogoliubov transformation, $\Psi_{\mathbf{k}}=\hat{\mathcal{U}}_{\mathbf{k}}\Phi_{\mathbf{k}}$.
To ensure that the transformed operators satisfy the correct bosonic
commutation relations, it is convenient to introduce the Bogoliubov
metric: 
\begin{equation}
\hat{\eta}=\text{diag}(1,-1,1,-1,...)\label{eq_metric}
\end{equation}

Then, the generalized Bogoliubov transformation satisfies: 
\begin{align}
\hat{\mathcal{U}}_{\mathbf{k}}^{\dagger}\,\hat{\eta}\,\hat{\mathcal{U}}_{\mathbf{k}} & =\mathbb{\hat{\eta}}\nonumber \\
\hat{\mathcal{U}}_{\mathbf{k}}^{-1}\left(\hat{\eta}\,\hat{\mathcal{H}}_{\mathbf{k}}\right)\hat{\mathcal{U}}_{\mathbf{k}} & =\left(\mathcal{\hat{\eta}\,\hat{\mathcal{H}}_{\mathbf{k}}}\right)^{\text{diag}}\label{eq:bogo_criter}
\end{align}

The spin-wave modes are therefore the eigenvalues of $\hat{\eta}\,\hat{\mathcal{H}}_{\mathbf{k}}$.

\subsection{Stripe phase: spin-wave modes}

As discussed above, the stripe phase is the ground state of the model
(\ref{eq:heisenberg_model}) for $K>0$. The spin-wave dispersion
of the stripe phase was obtained previously in Refs. \cite{Carlson08,Applegate10,fang}
and here we rederive the results to compare them later with the orthomagnetic
case. For concreteness, we first consider the stripe phase with ordering
vector $\mathbf{Q}_{1}=\left(\pi,0\right)$. As shown in Fig.\ref{fig:spin_config}(a),
there are two spins per magnetic unit cell, whose HP operators we
denote by $a_{\mathbf{k}}^{(1)}$ and $a_{\mathbf{k}}^{(2)}$. Note
that, with respect to the spin coordinate system defined on site 1,
the spin on site 2 is rotated by $180^{\circ}$, yielding: 
\begin{equation}
\mathbf{S}^{(2)}=\left(-S_{x}^{(2)},S_{y}^{(2)},-S_{z}^{(2)}\right)_{(1)}\label{eq_S_stripe}
\end{equation}

Using the Holstein-Primakoff transformation defined in Eq. \ref{eq:psi},
we find that the large-$S$ Hamiltonian is given by:

\begin{equation}
\hat{\mathcal{H}}_{\mathbf{k}}=\left(\begin{array}{cccc}
\epsilon_{\mathbf{k}} & 0 & 0 & \Delta_{\mathbf{k}}\\
0 & \epsilon_{\mathbf{k}} & \Delta_{\mathbf{k}} & 0\\
0 & \Delta_{\mathbf{k}} & \epsilon_{\mathbf{k}} & 0\\
\Delta_{\mathbf{k}} & 0 & 0 & \epsilon_{\mathbf{k}}
\end{array}\right)\label{H_stripe}
\end{equation}
with: 
\begin{align}
\epsilon_{\mathbf{k}} & =2S\left[\left(J_{1}-2K\right)\cos k_{y}+2J_{2}+4K\right]\nonumber \\
\Delta_{\mathbf{k}} & =-2S\left(J_{1}+2K+2J_{2}\cos k_{y}\right)\cos k_{x}\label{aux_H_stripe}
\end{align}

The Hamiltonian is diagonalized via the Bogoliubov transformation
\begin{equation}
\hat{\mathcal{U}}_{\mathbf{k}}=\left(\begin{array}{cccc}
u_{\mathbf{k}} & 0 & 0 & v_{\mathbf{k}}\\
0 & u_{\mathbf{k}} & v_{\mathbf{k}} & 0\\
0 & v_{\mathbf{k}} & u_{\mathbf{k}} & 0\\
v_{\mathbf{k}} & 0 & 0 & u_{\mathbf{k}}
\end{array}\right)\label{U_stripe}
\end{equation}
with:

\begin{equation}
u_{\mathbf{k}}^{2}=\frac{1}{2}\left(1+\frac{\epsilon_{\mathbf{k}}}{\omega_{\mathbf{k}}}\right);\ v_{\mathbf{k}}^{2}=\frac{1}{2}\left(-1+\frac{\epsilon_{\mathbf{k}}}{\omega_{\mathbf{k}}}\right);\ u_{\mathbf{k}}v_{\mathbf{k}}=-\frac{1}{2}\frac{\Delta_{\mathbf{k}}}{\omega_{\mathbf{k}}}\label{aux_U_stripe}
\end{equation}
yielding the doubly-degenerate eigenmode (i.e. spin-wave mode) of
the bosonic system:

\begin{equation}
\omega_{\mathbf{k}}=\sqrt{\epsilon_{\mathbf{k}}^{2}-\Delta_{\mathbf{k}}^{2}}\label{egv_stripe}
\end{equation}

\begin{figure}
\includegraphics[width=0.9\linewidth]{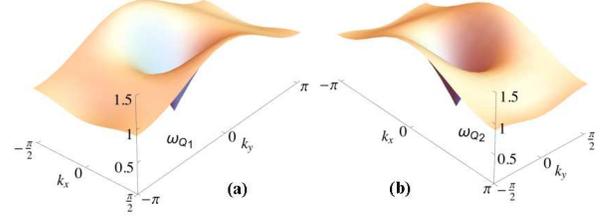}\protect\protect\protect\protect\caption{\label{fig:disp-stripe}Spin-wave dispersions for the stripe phase
with ordering vector $\mathbf{Q}_{1}=\left(\pi,0\right)$ (a) and
$\mathbf{Q}_{2}=\left(0,\pi\right)$ (b). Both are doubly degenerate
modes. The energies are in units of $4J_{2}S$ and the parameters
used are $J_{1}=0.8J_{2}$ and $K=0.1J_{2}$. }
\end{figure}

The fact that there are two degenerate spin-wave modes for the stripe
state is a consequence of the fact that $\omega_{\mathbf{k}+\mathbf{Q}_{1}}=\omega_{\mathbf{k}}$
and also of the collinear configuration of the spins. The spin-wave
dispersion of the stripe phase with ordering vector $\mathbf{Q}_{2}=\left(0,\pi\right)$
can be calculated in the same way, yielding $\omega_{\mathbf{Q}_{2}}\left(k_{x},k_{y}\right)=\omega_{\mathbf{Q}_{1}}\left(k_{y},-k_{x}\right)$,
as expected. In Fig. \ref{fig:disp-stripe}, we show the dispersion
of the spin waves (\ref{egv_stripe}) for the two types of stripe
orders in their respective \emph{magnetic} Brillouin zones. The results
obtained here are in agreement with those obtained previously elsewhere
\cite{fang}.

\subsection{Orthomagnetic phase: spin-wave modes}

The orthomagnetic phase becomes the ground state of Eq. (\ref{eq:heisenberg_model})
for $K<0$. As shown in Fig.\ref{fig:spin_config}(c), there are four
spins per magnetic unit cell, giving rise to the HP operators $a_{\mathbf{k}}^{(1)}$,
$a_{\mathbf{k}}^{(2)}$, $a_{\mathbf{k}}^{(3)}$, and $a_{\mathbf{k}}^{(4)}$.
Because the spins on sites 2, 3, 4 correspond respectively to rotations
of $90^{\circ}$, $180^{\circ}$, and $270^{\circ}$ relative to the
spin on site 1, we define the local spin coordinate systems: 
\begin{align}
\mathbf{S}^{(2)} & =(S_{z}^{(2)},S_{y}^{(2)},-S_{x}^{(2)})_{(1)}\nonumber \\
\mathbf{S}^{(3)} & =(-S_{z}^{(3)},S_{y}^{(3)},S_{x}^{(3)})_{(1)}\nonumber \\
\mathbf{S}^{(4)} & =(-S_{x}^{(4)},S_{y}^{(4)},-S_{z}^{(4)})_{(1)}\label{S_ortho}
\end{align}

Introducing $\Psi_{\mathbf{k}}^{\dagger}$ as defined in Eq. (\ref{eq:psi})
and substituting in the Hamiltonian, we obtain in the large-$S$ limit:
\begin{equation}
\hat{\mathcal{H}}_{\mathbf{k}}=\left(\begin{array}{cccc}
\hat{A}_{\mathbf{k}} & \hat{B}_{\mathbf{k}} & \hat{C}_{\mathbf{k}} & \hat{D}_{\mathbf{k}}\\
\hat{B}_{\mathbf{k}} & \hat{A}_{\mathbf{k}} & \hat{D}_{\mathbf{k}} & \hat{C}_{\mathbf{k}}\\
\hat{C}_{\mathbf{k}} & \hat{D}_{\mathbf{k}} & \hat{A}_{\mathbf{k}} & \hat{B}_{\mathbf{k}}\\
\hat{D}_{\mathbf{k}} & \hat{C}_{\mathbf{k}} & \hat{B}_{\mathbf{k}} & \hat{A}_{\mathbf{k}}
\end{array}\right)\label{H_ortho}
\end{equation}
where we defined four $2\times2$ matrices, $\hat{A}_{\mathbf{k}}$,
$\hat{B}_{\mathbf{k}}$, $\hat{C}_{\mathbf{k}}$, and $\hat{D}_{\mathbf{k}}$
of the form:

\begin{equation}
\hat{O}_{\mathbf{k}}=\left(\begin{array}{cc}
O_{1,\mathbf{k}} & O_{2,\mathbf{k}}\\
O_{2,\mathbf{k}} & O_{1,\mathbf{k}}
\end{array}\right)\label{aux_O}
\end{equation}
with the matrix elements:

\begin{flalign}
A_{1,\mathbf{k}} & =4S\left(J_{2}-K\right),\ A_{2,\mathbf{k}}=-4SK\nonumber \\
B_{1,\mathbf{k}} & =S\left(J_{1}+2K\right)\cos k_{x},\ B_{2,\mathbf{k}}=-S\left(J_{1}-2K\right)\cos k_{x}\nonumber \\
C_{1,\mathbf{k}} & =S\left(J_{1}+2K\right)\cos k_{y},\ C_{2,\mathbf{k}}=-S\left(J_{1}-2K\right)\cos k_{y}\nonumber \\
D_{1,\mathbf{k}} & =0,\ D_{2,\mathbf{k}}=-4SJ_{2}\cos k_{x}\cos k_{y}\label{aux_H_ortho}
\end{flalign}

The generalized Bogoliubov transformation is given by:

\begin{equation}
\hat{\mathcal{U}}_{\mathbf{k}}=\left(\begin{array}{cccc}
\hat{X}_{\mathbf{k}} & \hat{Y}_{\mathbf{k}} & \hat{Z}_{\mathbf{k}} & \hat{W}_{\mathbf{k}}\\
\hat{X}_{\mathbf{k}} & -\hat{Y}_{\mathbf{k}} & \hat{Z}_{\mathbf{k}} & -\hat{W}_{\mathbf{k}}\\
\hat{X}_{\mathbf{k}} & \hat{Y}_{\mathbf{k}} & -\hat{Z}_{\mathbf{k}} & -\hat{W}_{\mathbf{k}}\\
\hat{X}_{\mathbf{k}} & -\hat{Y}_{\mathbf{k}} & -\hat{Z}_{\mathbf{k}} & \hat{W}_{\mathbf{k}}
\end{array}\right)\label{U_ortho}
\end{equation}
where the four $2\times2$ matrices, $\hat{X}_{\mathbf{k}}$, $\hat{Y}_{\mathbf{k}}$,
$\hat{Z}_{\mathbf{k}}$, and $\hat{W}_{\mathbf{k}}$ are also of the
form (\ref{aux_O}). For $\hat{X}_{\mathbf{k}}$, the matrix elements
are given by:

\begin{align}
X_{1,\mathbf{k}}^{2} & =\frac{1}{8}\left(1+\frac{\epsilon_{\mathbf{k}}}{\omega_{\mathbf{k}}}\right)\nonumber \\
X_{2,\mathbf{k}}^{2} & =\frac{1}{8}\left(-1+\frac{\epsilon_{\mathbf{k}}}{\omega_{\mathbf{k}}}\right)\nonumber \\
X_{1,\mathbf{k}}X_{2,\mathbf{k}} & =-\frac{1}{8}\frac{\Delta_{\mathbf{k}}}{\omega_{\mathbf{k}}}\label{aux_U_ortho}
\end{align}
with: 
\begin{align}
\epsilon_{\mathbf{k}} & =A_{1,\mathbf{k}}+B_{1,\mathbf{k}}+C_{1,\mathbf{k}}+D_{1,\mathbf{k}}\nonumber \\
\Delta_{\mathbf{k}} & =A_{2,\mathbf{k}}+B_{2,\mathbf{k}}+C_{2,\mathbf{k}}+D_{2,\mathbf{k}}\label{e_delta_ortho}
\end{align}
and the spin-wave dispersions:

\begin{equation}
\omega_{\mathbf{k}}=\sqrt{\epsilon_{\mathbf{k}}^{2}-\Delta_{\mathbf{k}}^{2}}\label{egv_ortho}
\end{equation}

For the other matrix elements, we find:

\begin{align}
Y_{i,\mathbf{k}} & =X_{i,\mathbf{k}+\mathbf{Q}_{1}}\nonumber \\
Z_{i,\mathbf{k}} & =X_{i,\mathbf{k}+\mathbf{Q}_{2}}\nonumber \\
W_{i,\mathbf{k}} & =X_{i,\mathbf{k}+\mathbf{Q}_{1}+\mathbf{Q}_{2}}\label{aux_Y_Z_W}
\end{align}

\begin{figure}
\includegraphics[width=0.9\linewidth]{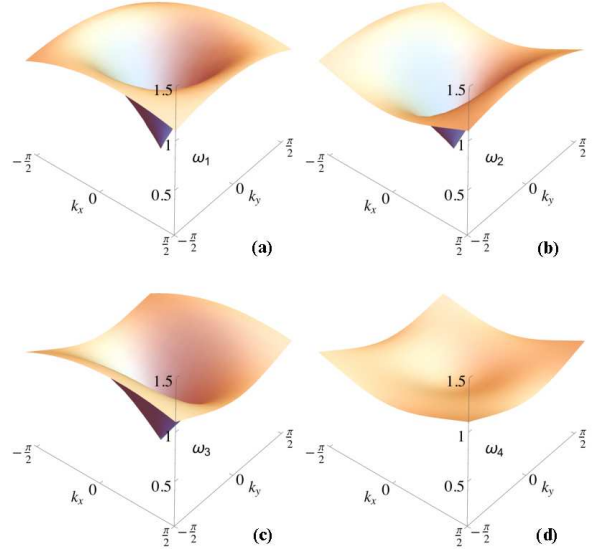}

\protect\protect\protect\protect\caption{\label{fig:disp-ortho}The spin-wave dispersions of the non-collinear
orthomagnetic order in the magnetic Brillouin zone. The four dispersions
are linked by a shift of the momentum coordinate system by the orthomagnetic
ordering vectors. Here, the energies are in units of $4J_{2}S$ and
the parameters are $J_{1}=0.8J_{2}$ and $K=-0.1J_{2}$. }
\end{figure}

Therefore, there are four non-degenerate spin-wave dispersions of
the bosonic system:

\begin{equation}
\omega_{1\mathbf{k}}=\omega_{\mathbf{k}};\ \omega_{2\mathbf{k}}=\omega_{\mathbf{k}+\mathbf{Q}_{1}};\ \omega_{3\mathbf{k}}=\omega_{\mathbf{k}+\mathbf{Q}_{2}};\ \omega_{4\mathbf{k}}=\omega_{\mathbf{k}+\mathbf{Q}_{1}+\mathbf{Q}_{2}}\label{spin_waves_ortho}
\end{equation}
with $\omega_{\mathbf{k}}$ given in Eq. (\ref{egv_ortho}). These
four spin-wave dispersions are shifted with respect to each other
by the ordering vectors of the orthomagnetic phase, corresponding
to in-phase or out-of-phase combinations of the four HP bosons. All
of them are shown in Fig. \ref{fig:disp-ortho} in the \emph{magnetic}
unit cell of the orthomagnetic phase. We note that while $\omega_{1}$,
$\omega_{2}$, and $\omega_{3}$ display gapless modes, corresponding
to three Goldstone modes, the $\omega_{4}$ spin-wave dispersion is
gapped. The fact that there are three Goldstone modes is a consequence
of the non-collinear magnetic configuration of the orthomagnetic phase,
which breaks completely all the spin-rotational symmetries of the
system.

\subsection{Dynamic structure factors of the stripe and orthomagnetic phases}

Having established the nature of the spin-wave modes in the stripe
and orthomagnetic phases, we now proceed to compute the spin-spin
correlation function $\mathcal{S}_{\mu\nu}$ in the non-magnetic unit
cell, which can be measured by neutron scattering. We have \cite{haraldsen,Carlson04}:
\begin{equation}
\mathcal{S}_{\mu\nu}\left(\mathbf{k},\omega\right)=\int_{-\infty}^{\infty}\frac{\mathrm{d}t}{2\pi}\ \mathrm{e}^{-i\omega t}\langle S_{\mathbf{k}}^{\mu}(0)S_{-\mathbf{k}}^{\nu}(t)\rangle\label{eq:spin-corr}
\end{equation}
where $\mu\nu=x,y,z$ refer to the spin components and $\mathbf{S}_{\mathbf{k}}\equiv\sum_{r}\mathbf{S}_{r,\mathbf{k}}$
is the sum over all the $r$ spins in the magnetic unit cell. Here,
the spin coordinate system is defined globally with respect to the
neutron polarization, in contrast to the local coordinate system introduced
in the previous subsection. For concreteness, hereafter we assume
the incoming neutron to be polarized parallel to the spin on site
$1$, i.e. parallel to the $\hat{\mathbf{z}}$ axis. Computation of
Eq. (\ref{eq:spin-corr}) is straightforward with the aid of the HP
bosons and the Bogoliubov transformation defined in the previous subsection.
Denoting by $\tilde{a}_{\mathbf{k}}^{(r)}$ the Bogoliubov-transformed
bosonic operators, the only non-zero terms, at $T=0$, are those of
the form:

\begin{equation}
\int_{-\infty}^{\infty}\frac{\mathrm{d}t}{2\pi}\ \mathrm{e}^{-i\omega t}\left\langle \tilde{a}_{\mathbf{k}}^{(r)}\left(0\right)\tilde{a}_{\mathbf{k}}^{\dagger(r)}\left(t\right)\right\rangle =\delta\left(\omega-\omega_{r,\mathbf{k}}\right)\label{aux_spin-corr}
\end{equation}

\begin{figure}
\includegraphics[width=0.8\linewidth]{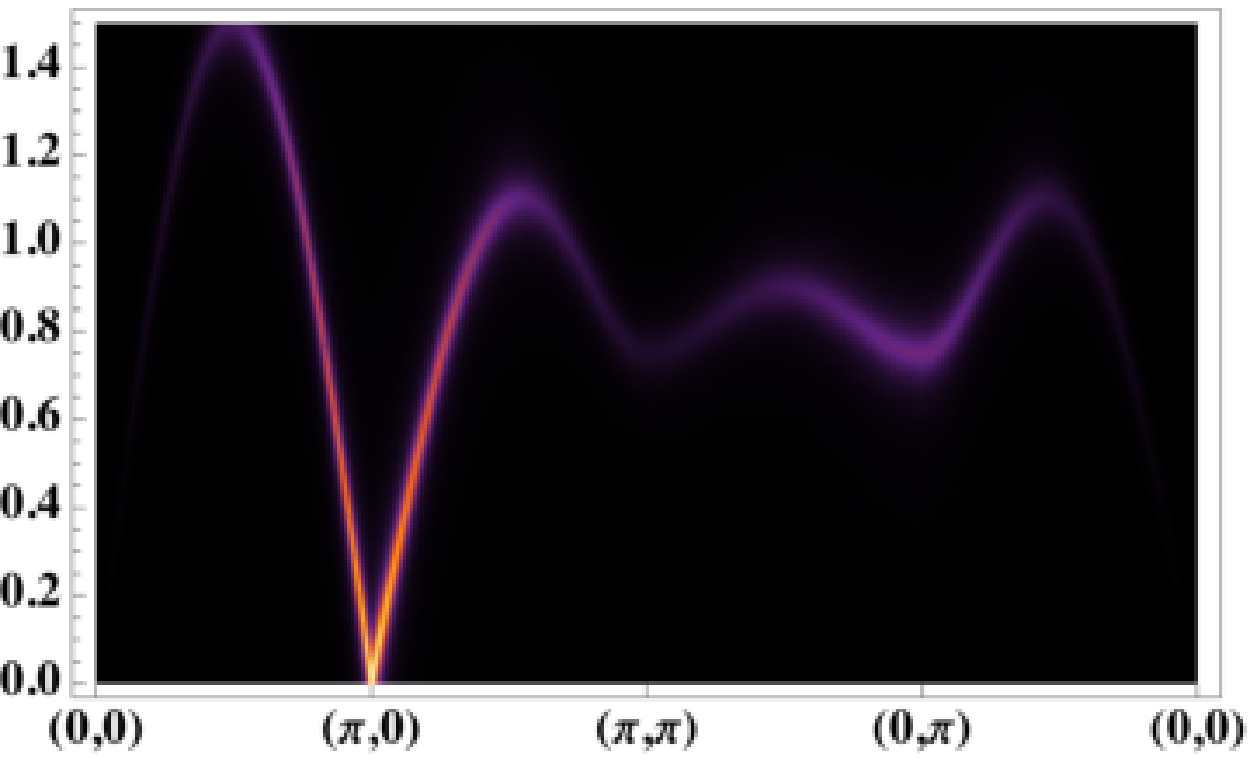}

\includegraphics[width=0.8\linewidth]{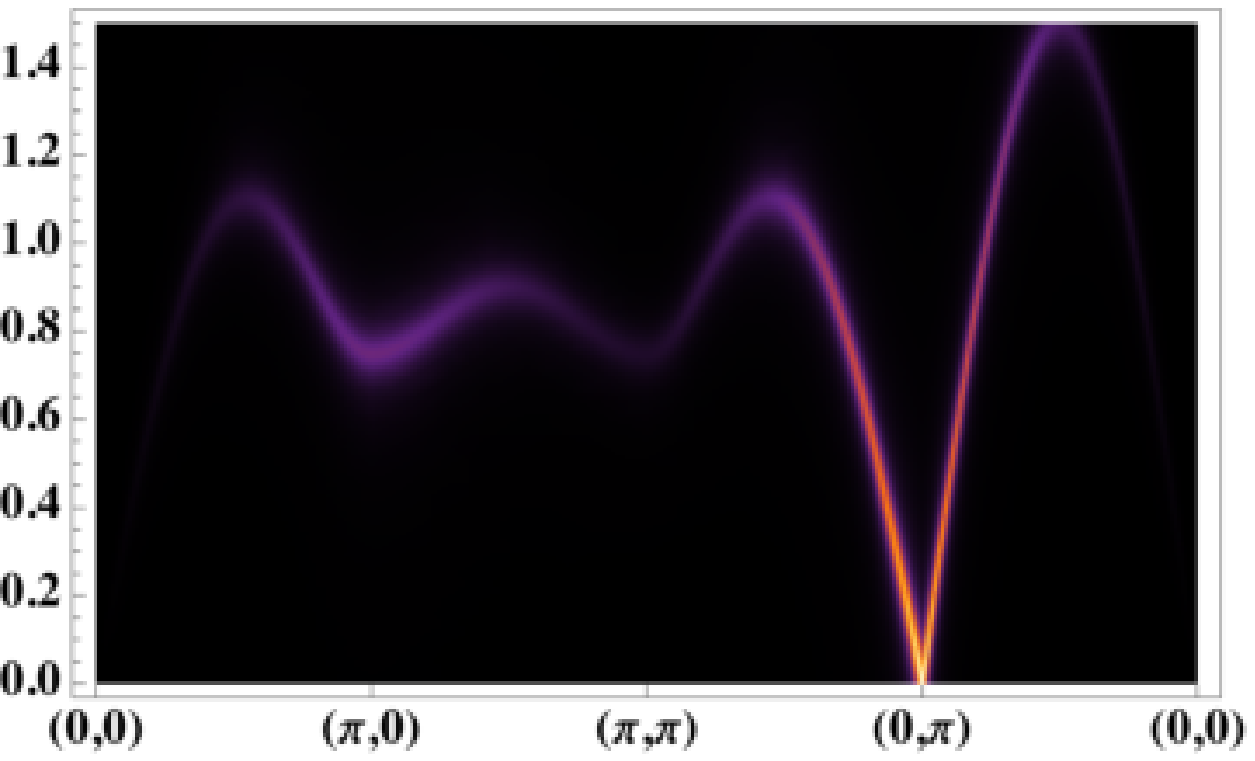}

\includegraphics[width=0.8\linewidth]{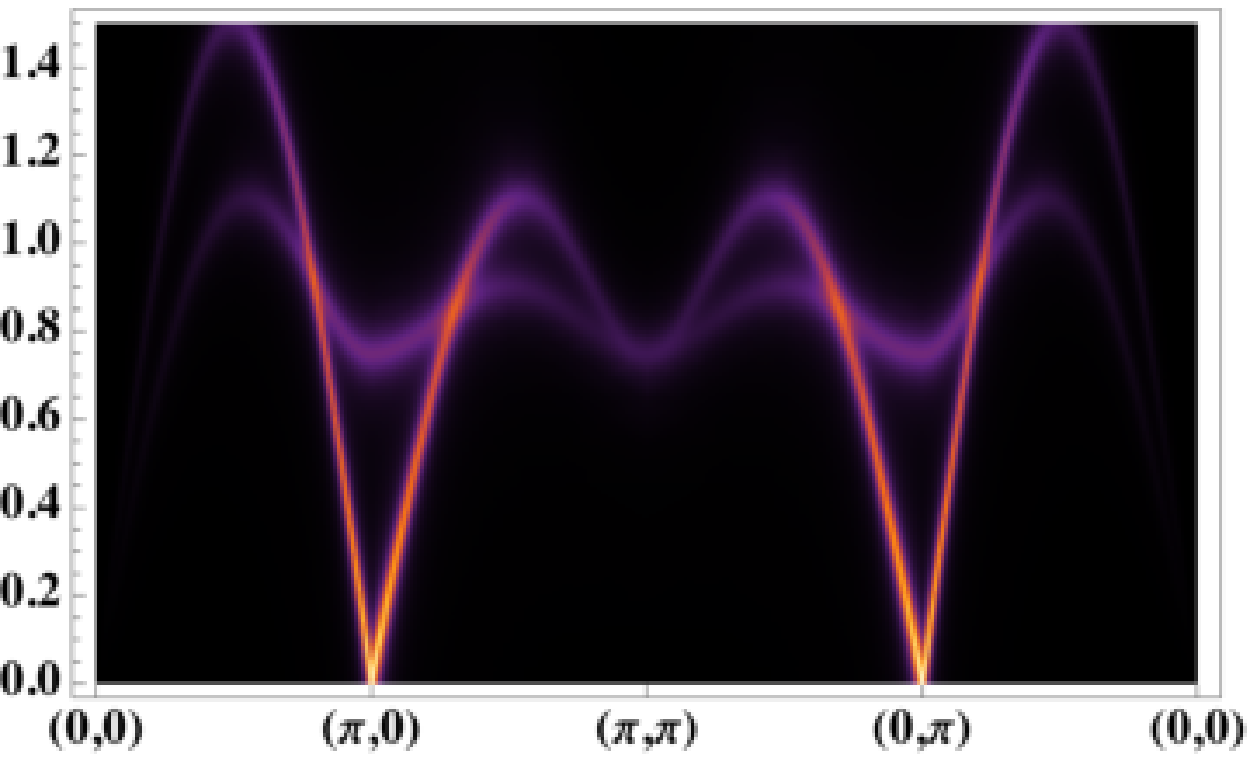}

\protect\protect\protect\protect\caption{\label{fig:stripe_struct}The total spin-spin structure factor $\mathcal{S}=\sum_{i}\mathcal{S}_{ii}=2S_{xx}$
for the $\mathbf{Q}_{1}$ (top panel) and $\mathbf{Q}_{2}$ (mid panel)
stripe phases. Bottom panel is the structure factor assuming equal
domains of $\mathbf{Q}_{1}$ and $\mathbf{Q}_{2}$ stripes. The vertical
axis is the energy measured in units of $4J_{2}S$, whereas the horizontal
axis displays momentum cuts in the Fe-square-lattice Brillouin zone.
The intensity is highest at the ordering vectors $\mathbf{Q}_{1}$
and $\mathbf{Q}_{2}$. The parameters used here are $J_{1}=0.8J_{2}$
and $K=0.1J_{2}$.}
\end{figure}

We first consider the stripe phase with the two possible ordering
vectors $\mathbf{Q}_{1}=\left(\pi,0\right)$ and $\mathbf{Q}_{2}=\left(0,\pi\right)$.
We find that only the transverse components $\mathcal{S}_{xx}=\mathcal{S}_{yy}$
are non-zero, i.e. the longitudinal component $S_{zz}$ and the off-diagonal
components $\mathcal{S}_{i\neq j}$ do not acquire spin-wave contributions.
We obtain:

\begin{equation}
\mathcal{S}_{xx}^{\mathbf{Q}_{i}}\left(\mathbf{k},\omega\right)=2S\left(u_{\mathbf{k}}-v_{\mathbf{k}}\right)^{2}\delta\left(\omega-\omega_{\mathbf{k}}^{\mathbf{Q}_{i}}\right)\label{S_stripe}
\end{equation}
where $u_{\mathbf{k}},v_{\mathbf{k}}$ are given by Eq. (\ref{aux_U_stripe})
and $\omega_{\left(k_{x},k_{y}\right)}^{\mathbf{Q}_{2}}=\omega_{\left(-k_{y},k_{x}\right)}^{\mathbf{Q}_{1}}$,
with $\omega_{\mathbf{k}}^{\mathbf{Q}_{1}}$ given by Eq. (\ref{egv_stripe}).
The total spin-spin correlation function, $\mathcal{S}=\sum_{i}\mathcal{S}_{ii}$
is then simply $\mathcal{S}\left(\mathbf{k},\omega\right)=2\mathcal{S}_{xx}\left(\mathbf{k},\omega\right)$.
In Fig. \ref{fig:stripe_struct}, we plot $\mathcal{S}(\mathbf{k},\omega)$
for both the $\mathbf{Q}_{1}$ and $\mathbf{Q}_{2}$ stripe phases
separately, as well as for a system containing equal domains of $\mathbf{Q}_{1}$
and $\mathbf{Q}_{2}$:

\begin{equation}
\mathcal{S}_{\mathrm{domain}}\left(\mathbf{k},\omega\right)=\frac{1}{2}\mathcal{S}^{\mathbf{Q}_{1}}\left(\mathbf{k},\omega\right)+\frac{1}{2}\mathcal{S}^{\mathbf{Q}_{2}}\left(\mathbf{k},\omega\right)\label{domains}
\end{equation}

The latter is the case relevant for the real systems, since twin domains
are always formed in the iron pnictides. In all the plots, the delta
function is replaced by a Lorentzian with width $\gamma=0.05$ in
units of $2J_{2}S$. From the figure, we see that the system with
twin domains display anisotropic spin-wave branches emerging from
the ordering vectors $\mathbf{Q}_{1}=\left(\pi,0\right)$ and $\mathbf{Q}_{2}=\left(0,\pi\right)$,
as expected. In all cases, the structure factor vanishes at center
of the Brillouin zone, but diverges at the ordering vectors $\mathbf{Q}_{i}$.
Therefore, expanding the spin-wave dispersion around the ordering
vector $\mathbf{Q}_{i}$ yields ($\theta$ denote the polar angle
between $\mathbf{k}$ and $\hat{\mathbf{k}}_{x}$): 
\begin{equation}
\omega_{\mathbf{k}+\mathbf{Q}_{i}}\approx4S\left|\mathbf{k}\right|\sqrt{\left(J_{2}+K+\frac{J_{1}}{2}\right)\left(J_{2}+K\pm\frac{J_{1}}{2}\cos2\theta\right)}\label{SW_stripe_expansion}
\end{equation}
which is anisotropic along the $k_{x}$ and $k_{y}$ axis, as expected.
In the previous expression, the upper (lower) sign refers to $\mathbf{Q}_{1}$
($\mathbf{Q}_{2}$).

\begin{figure}
\includegraphics[width=0.8\linewidth]{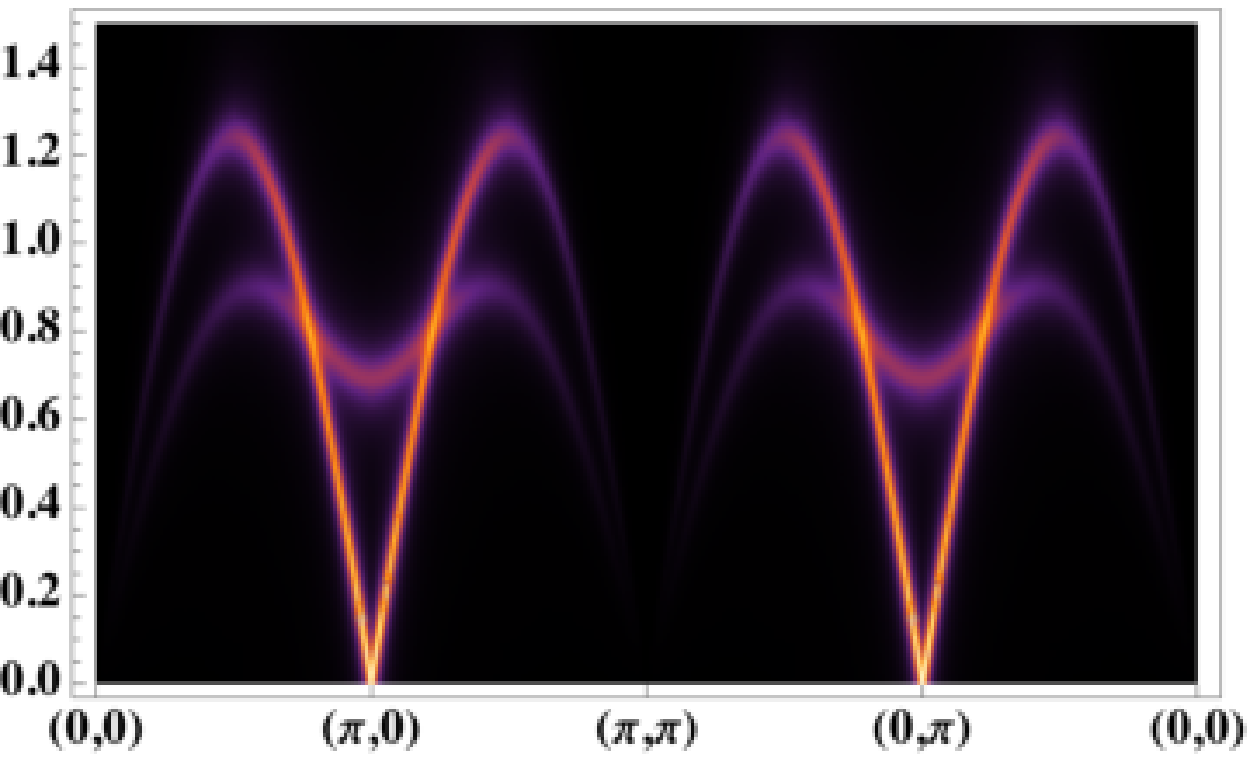}

\includegraphics[width=0.8\linewidth]{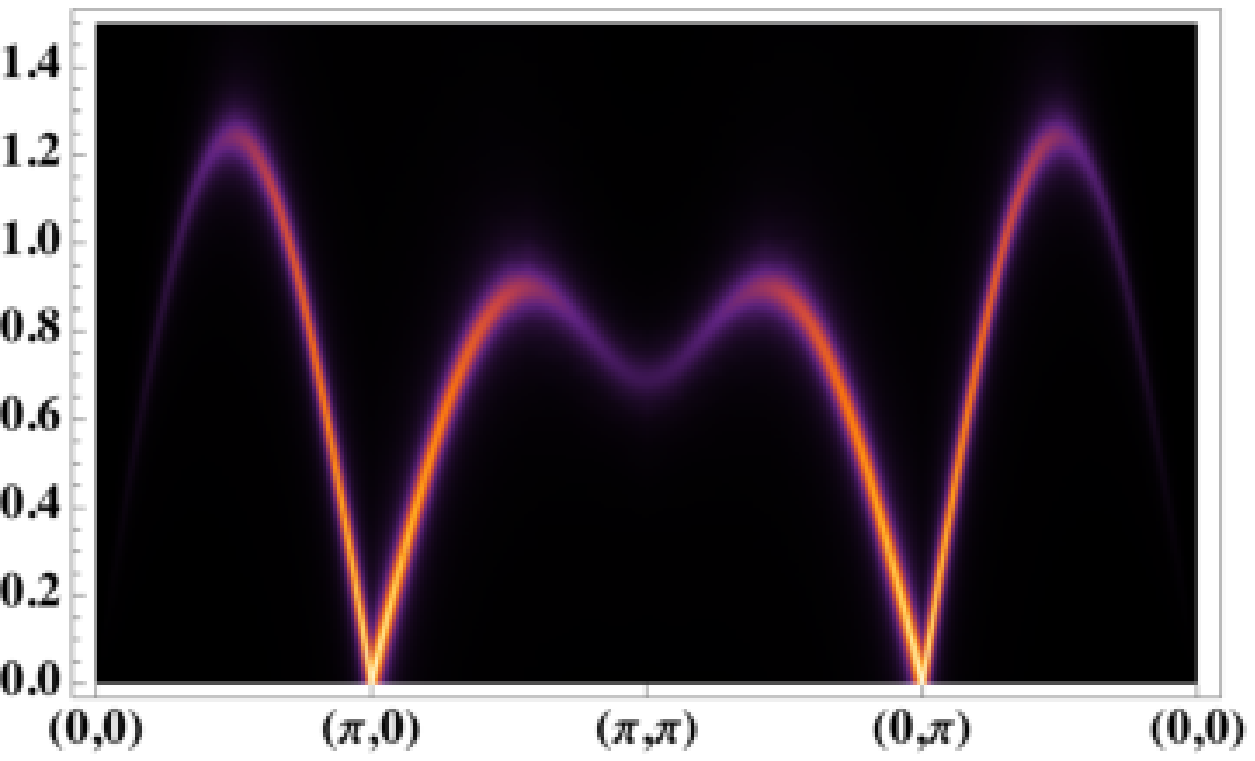}

\includegraphics[width=0.8\linewidth]{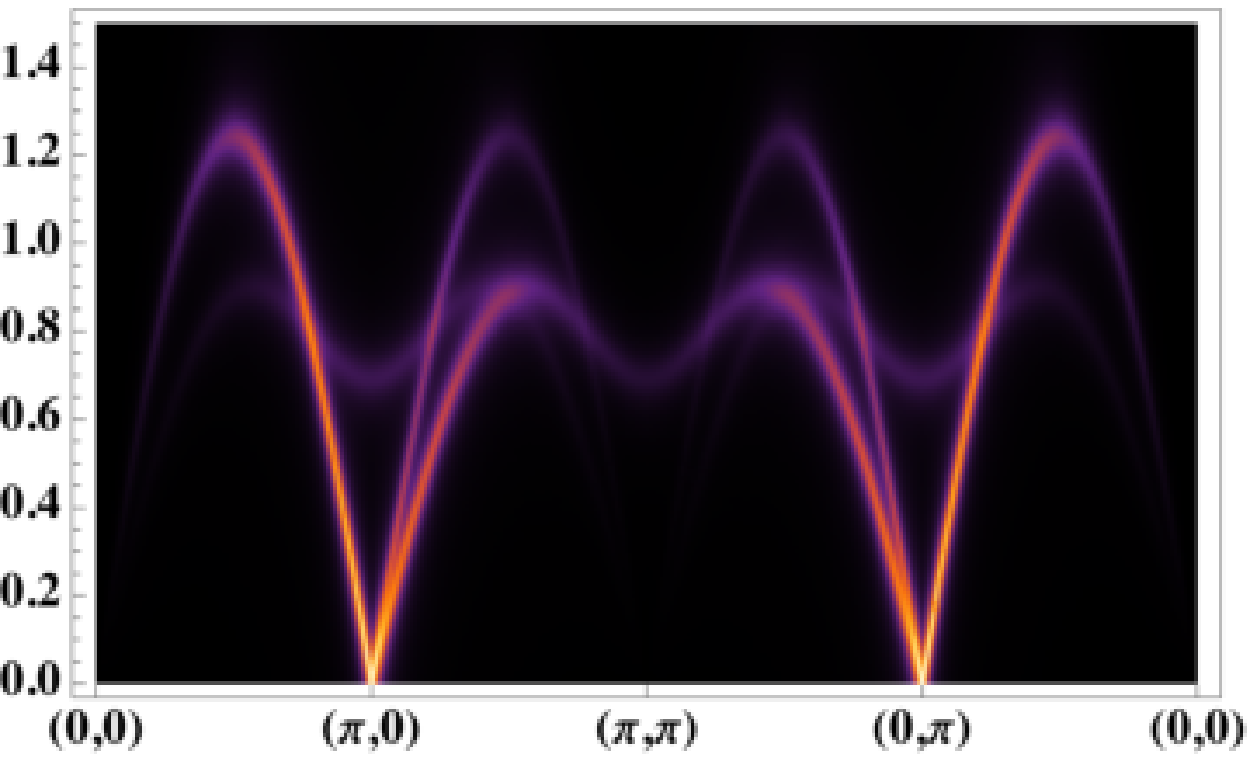}

\protect\protect\protect\protect\caption{\label{fig:ortho_struct}The structure factors $\mathcal{S}_{xx}=\mathcal{S}_{zz}$
(top panel), $\mathcal{S}_{yy}$ (mid panel) and $\mathcal{S}=\sum_{i}\mathcal{S}_{ii}$
(bottom panel) for the orthomagnetic phase. The vertical axis is the
energy measured in units of $4J_{2}S$, whereas the horizontal axis
displays momentum cuts in the Fe-square-lattice Brillouin zone. The
parameters used here are $J_{1}=0.8J_{2}$ and $K=-0.1J_{2}$. Two
gapless spin-wave branches emerge from the ordering vectors $\mathbf{Q}_{1}=\left(\pi,0\right)$
and $\mathbf{Q}_{2}=\left(0,\pi\right)$, in sharp contrast to the
case of domains of stripes shown in Fig. \ref{fig:stripe_struct}. }
\end{figure}

For the orthomagnetic phase, we find that all diagonal components
$\mathcal{S}_{ii}$ acquire spin-wave contributions. This is expected
since the magnetic configuration is non-collinear (see Fig. \ref{fig:spin_config}),
implying that all directions are ``transverse'' with respect to
the sublattice $1,3$ and/or the sublattice $2,4$. In particular,
we find: 
\begin{align}
\mathcal{S}_{xx}\left(\mathbf{k},\omega\right)=\mathcal{S}_{zz}\left(\mathbf{k},\omega\right) & =4S\left(Y_{1,\mathbf{k}}+Y_{2,\mathbf{k}}\right)^{2}\delta\left(\omega-\omega_{2,\mathbf{k}}\right)\nonumber \\
 & +4S\left(Z_{1,\mathbf{k}}+Z_{2,\mathbf{k}}\right)^{2}\delta\left(\omega-\omega_{3,\mathbf{k}}\right)\nonumber \\
\mathcal{S}_{yy}\left(\mathbf{k},\omega\right) & =16S\left(X_{1,\mathbf{k}}-X_{2,\mathbf{k}}\right)^{2}\delta\left(\omega-\omega_{1,\mathbf{k}}\right)\label{S_ii_ortho}
\end{align}
with the Bogoliubov transformation parameters and spin-wave modes
defined in Eqs. (\ref{aux_U_ortho}) and (\ref{spin_waves_ortho}).
In Fig. \ref{fig:ortho_struct}, we plot these diagonal components
as well as the total structure factor $\mathcal{S}=\sum_{i}\mathcal{S}_{ii}$.
In the latter, we observe two spin-wave branches emerging from the
ordering vectors $\mathbf{Q}_{1}=\left(\pi,0\right)$ and $\mathbf{Q}_{2}=\left(0,\pi\right)$,
in sharp contrast to the case of domains of stripes, where only one
spin-wave branch emerges from each ordering vector (see Fig. \ref{fig:stripe_struct}).
We note that, once again, the structure factor vanishes at the center
of the Brillouin zone and diverges at the magnetic ordering vectors
$\mathbf{Q}_{1}=\left(\pi,0\right)$ and $\mathbf{Q}_{2}=\left(0,\pi\right)$.
Expanding the dispersions near them, we find (recall that $K<0$):

\begin{align}
\omega_{1,\mathbf{k}+\mathbf{Q}_{1}} & \approx4S\left|\mathbf{k}\right|\sqrt{\left(J_{2}-K\right)\left(J_{2}+\frac{J_{1}}{2}\cos2\theta\right)}\nonumber \\
\omega_{2,\mathbf{k}+\mathbf{Q}_{1}} & \approx4S\left|\mathbf{k}\right|\sqrt{\left(J_{2}-K\right)\left(J_{2}+\frac{J_{1}}{2}\right)}\nonumber \\
\omega_{3,\mathbf{k}+\mathbf{Q}_{1}} & \approx8S\sqrt{\left(2J_{2}-J_{1}\right)\left(-K\right)}\label{eq:SW_ortho_expansion_1}
\end{align}
as well as:

\begin{align}
\omega_{1,\mathbf{k}+\mathbf{Q}_{2}} & \approx4S\left|\mathbf{k}\right|\sqrt{\left(J_{2}-K\right)\left(J_{2}-\frac{J_{1}}{2}\cos2\theta\right)}\nonumber \\
\omega_{2,\mathbf{k}+\mathbf{Q}_{2}} & \approx8S\sqrt{\left(2J_{2}-J_{1}\right)\left(-K\right)}\nonumber \\
\omega_{3,\mathbf{k}+\mathbf{Q}_{2}} & \approx4S\left|\mathbf{k}\right|\sqrt{\left(J_{2}-K\right)\left(J_{2}+\frac{J_{1}}{2}\right)}\label{eq:SW_ortho_expansion_2}
\end{align}

Therefore, we obtain two gapless spin-wave branches emerging from
each ordering vector, as shown in Fig. \ref{fig:ortho_struct}, as
well as one gapped spin-wave dispersion. As expected, tetragonal symmetry
is preserved by these dispersions. Interestingly, along the direction
parallel to the $\mathbf{Q}_{i}$ vector, the two spin-wave velocities
are equal, whereas along the direction perpendicular to the $\mathbf{Q}_{i}$
vector, they are different. In the latter case, their ratio is given
by:

\begin{equation}
\frac{\mathrm{c}_{1,\perp}}{\mathrm{c}_{2,\perp}}=\sqrt{\frac{2J_{2}+J_{1}}{2J_{2}-J_{1}}}\label{SW_velocities}
\end{equation}
where the $\perp$ sign indicates that the spin-wave velocity is measured
relative to the direction perpendicular to the ordering vector $\mathbf{Q}_{i}$.
Interestingly, this ratio does not depend on the biquadratic coupling
$K$. These qualitative features, in principle, allow one to experimentally
distinguish, in an unambiguous way, whether the magnetic ground state
is stripe or orthomagnetic. Note that, in the orthomagnetic phase,
no spin-wave modes emerge from $\mathbf{Q}_{1}+\mathbf{Q}_{2}=\left(\pi,\pi\right)$.

Continuing the investigation of the orthomagnetic phase, we find that
the spin-waves also contribute to the off-diagonal component:

\begin{align}
\mathcal{S}_{xz}\left(\mathbf{k},\omega\right) & =4S\left(Y_{1,\mathbf{k}}+Y_{2,\mathbf{k}}\right)^{2}\delta\left(\omega-\omega_{2,\mathbf{k}}\right)\nonumber \\
 & -4S\left(Z_{1,\mathbf{k}}+Z_{2,\mathbf{k}}\right)^{2}\delta\left(\omega-\omega_{3,\mathbf{k}}\right)\label{S_ij_ortho}
\end{align}
providing another criterion to distinguish experimentally the orthomagnetic
and stripe phases via polarized neutron scattering.

In principle, the structure factor tensor of the orthomagnetic phase
can be brought in a diagonal form if the neutron is polarized along
$\tilde{\mathbf{z}}=\left(\mathbf{x}+\mathbf{z}\right)/\sqrt{2}$,
instead of parallel to the spin on site 1. In this new coordinate
system, each of the three gapless spin-wave dispersions contribute
only to one of the diagonal components, and we find:

\begin{align}
\mathcal{\tilde{S}}_{zz}\left(\mathbf{k},\omega\right) & =8S\left(Y_{1,\mathbf{k}}+Y_{2,\mathbf{k}}\right)^{2}\delta\left(\omega-\omega_{2,\mathbf{k}}\right)\nonumber \\
\mathcal{\tilde{S}}_{xx}\left(\mathbf{k},\omega\right) & =8S\left(Z_{1,\mathbf{k}}+Z_{2,\mathbf{k}}\right)^{2}\delta\left(\omega-\omega_{3,\mathbf{k}}\right)\label{S_tilde_ortho}
\end{align}
as well as $\tilde{\mathcal{S}}_{yy}=\mathcal{S}_{yy}$.

\section{Discussion and conclusions}

We investigated in details under what conditions the orthomagnetic
state, which displays a non-collinear double-\textbf{$\mathbf{Q}$
}tetragonal magnetic structure, becomes the magnetic ground state
of the iron pnictides within a microscopic itinerant model. We found
that large deviations from perfect nesting favor a tetragonal magnetic
state, but do not select between the non-collinear and non-uniform
configurations -- Figs. \ref{fig:spin_config}(c) and (d), respectively.
Instead, the non-collinear order is favored by the residual electronic
interactions that do not participate in the SDW -- in agreement with
the results found in Ref. \cite{Eremin10} -- whereas the non-uniform
state is favored by coupling to soft Neel-like fluctuations, as discussed
by two of us in Ref. \cite{xiaoyu}. Our investigation is complementary
to previous works reporting that different regions of the large parameters
space of the iron-based superconductors may display magnetic ground
states that do not break tetragonal symmetry \cite{Lorenzana08,Eremin10,Brydon11,xiaoyu}.
In particular, Ref. \cite{avci} showed that the same three-band model
studied in Section II accounts for a transition inside the magnetic
stripe state to the $C_{4}$-magnetic state. Our findings reveal that
the tetragonal magnetic state can also appear as the primary magnetic
instability of the system, without requiring pre-existing stripe order.

The significance of the observation of tetragonal magnetic states
in $\text{Ba}(\text{Fe}_{1-x}\text{Mn}_{x})_{2}\text{As}_{2}$, $(\text{Ba}_{1-x}\text{K}_{x})\text{Fe}_{2}\text{As}_{2}$,
and $(\text{Ba}_{1-x}\text{Na}_{x})\text{Fe}_{2}\text{As}_{2}$ relies
on its implication to the nature of the magnetism of these materials.
First, the existence of magnetic Bragg peaks at $\mathbf{Q}_{1}=\left(\pi,0\right)$
and $\mathbf{Q}_{2}=\left(0,\pi\right)$ in the absence of a splitting
of the lattice Bragg peaks implies that the tetragonal symmetry-breaking
is not a prerequisite for the formation of magnetic order, challenging
the point of view that ferro-orbital order is the leading normal-state
instability. Furthermore, because the non-collinear and non-uniform
magnetic states do not belong to the ground state manifold of the
$J_{1}$-$J_{2}$ model, the latter is likely not the most suitable
low-energy model to describe the magnetic properties of the iron-based
superconductors.

Of course, these statements rely on the confirmation that $\text{Ba}(\text{Fe}_{1-x}\text{Mn}_{x})_{2}\text{As}_{2}$,
$(\text{Ba}_{1-x}\text{K}_{x})\text{Fe}_{2}\text{As}_{2}$, and $(\text{Ba}_{1-x}\text{Na}_{x})\text{Fe}_{2}\text{As}_{2}$
do display tetragonal magnetic order. Up to now, the observations
have focused on the absence of detectable structural distortion, which
is usually large in most iron-based materials \cite{kim} due to their
sizable magneto-elastic coupling \cite{Patz14}. Nevertheless, given
the resolution limitations intrinsic to x-ray and neutron diffraction
probes \cite{inosov}, it is desirable to find other signatures of
these tetragonal magnetic states. Here, we have shown how qualitative
features in the spin-wave spectrum can unambiguously distinguish between
the stripe and orthomagnetic (non-collinear) phases. For instance,
while the latter displays two anisotropic gapless spin-wave branches
emerging from each of the ordering vectors $\mathbf{Q}_{1}=\left(\pi,0\right)$
and $\mathbf{Q}_{2}=\left(0,\pi\right)$, a system with domains of
the two distinct stripe states displays a single gapless spin-wave
branch emerging from each of them. Furthermore, only in the orthomagnetic
phase the spin waves can also be detected in off-diagonal components
of the spin-spin correlation function. These two distinguishing features
can in principle be probed by unpolarized and polarized neutron scattering
experiments, respectively. We have not discussed the spin-wave spectrum
of the non-uniform phase, which is beyond the scope of the current
paper. Yet, because this state is collinear, one does not expect the
appearance of additional Goldstone modes, as in the orthomagnetic
phase. Interesting features can appear at the ordering vector $\mathbf{Q}_{1}+\mathbf{Q}_{2}=\left(\pi,\pi\right)$
in the non-collinear phase due to the formation of a composite order
parameter, as discussed in Ref. \cite{xiaoyu}. It remains to be seen
how these tetragonal magnetic states affect the superconducting state.

Finally, we note that the results obtained here for the spin-wave
spectra of the stripe and orthomagnetic phases can also be useful
to determine the magnetic states of other compounds that display magnetic
Bragg peaks at $\mathbf{Q}_{1}=\left(\pi,0\right)$ and $\mathbf{Q}_{2}=\left(0,\pi\right)$
but no orthorhombic distortion. In general, without knowledge of the
size of the magneto-elastic coupling, it is difficult to establish
whether these observations are consistent with domains of stripes
or orthomagnetic order. A recent example is the compound GdRhIn$_{5}$,
which is related to the 115 family of heavy fermions \cite{granado}.
Although resonant x-ray scattering found evidence for magnetic order
at momenta $\mathbf{Q}_{1}$ and $\mathbf{Q}_{2}$, synchrotron x-rays
were unable to resolve a structural distortion. Furthermore, the magnetic
transition seems to be second-order, which is difficult to reconcile
with a simultaneous structural transition. An interesting alternative
would be the formation of an orthomagnetic state, which could be identified
by neutron scattering experiments deep inside the magnetically ordered
state.

\section{Acknowledgements}

We thank A. Böhmer, P. Canfield, M. Chan, A. Chubukov, I. Eremin,
A. Goldman, M. Greven, C. Meingast, R. McQueeney, R. Osborn, P. Pagliuso,
L. Taillefer, and G. Yu for fruitful discussions. This work was supported
by the U.S. Department of Energy under Award Number DE-SC0012336.

\appendix

\section{Contribution of the residual interactions to the free energy}

Here we show how to explicitly compute the contribution of the residual
interactions $U_{2}$, $U_{4}$, $U_{5}$, $U_{6}$, $U_{7}$, and
$U_{8}$ in Eq. (\ref{eq:int_hamltonian}) to the free energy. We
illustrate the procedure by considering the $U_{7}$ term, corresponding
to an exchange-like interaction between the electron pockets at $\mathbf{Q}_{1}$
and $\mathbf{Q}_{2}$. To lowest order in $U_{7}$, the contribution
to the free energy corresponds to the two Feynman diagrams shown in
Fig. \ref{fig_ee_interaction} in the main text. Because we are interested
in the uniform limit of the action, the momentum of the fields $\mathbf{M}_{1}$
and $\mathbf{M}_{2}$ is set to zero. Also, because we are approaching
the transition from the paramagnetic side, we ignore the corrections
to the electronic Green's functions due to the presence of SDW order.

Denoting the generalized momentum by $q$, the left diagram corresponds
to 
\begin{align}
U_{7}\left(M_{1}^{i}M_{2}^{l}M_{2}^{j}M_{1}^{k}\right)\left(\sigma_{\alpha\mu}^{i}\sigma_{\mu\alpha}^{l}\sigma_{\beta\nu}^{j}\sigma_{\nu\beta}^{k}\right)\nonumber \\
\left[G_{h}\left(q\right)G_{e_{1}}\left(q\right)G_{e_{2}}\left(q\right)G_{h}\left(p\right)G_{e_{1}}\left(p\right)G_{e_{2}}\left(p\right)\right]\label{A1}
\end{align}
where $i,j,k,l$ correspond to vector components and repeated index
are implicitly summed. The sum over Pauli matrices is equivalent to:

\begin{equation}
\text{Tr}\left\{ \sigma^{i}\sigma^{l}\right\} \text{Tr}\left\{ \sigma^{j}\sigma^{k}\right\} =4\delta^{il}\delta^{jk}\label{A2}
\end{equation}

As a result, this diagram gives the contribution 
\begin{equation}
4U_{7}\left(\sum G_{h}G_{e_{1}}G_{e_{2}}\right)^{2}\left(\mathbf{M}_{1}\cdot\mathbf{M}_{2}\right)^{2}\label{A3}
\end{equation}

The right diagram corresponds to:

\begin{align}
-U_{7}\left(M_{1}^{i}M_{2}^{l}M_{2}^{j}M_{1}^{k}\right)\left(\sigma_{\mu\beta}^{i}\sigma_{\beta\nu}^{l}\sigma_{\nu\alpha}^{j}\sigma_{\alpha\mu}^{k}\right)\nonumber \\
\left[G_{h}\left(q\right)G_{e_{1}}\left(q\right)G_{e_{2}}\left(q\right)G_{h}\left(p\right)G_{e_{1}}\left(p\right)G_{e_{2}}\left(p\right)\right]\label{A4}
\end{align}
where the minus sign comes from the closed fermionic loop. Using the
identity: 
\begin{equation}
\text{Tr\ensuremath{\left\{ \sigma^{i}\sigma^{l}\sigma^{j}\sigma^{k}\right\} }=2\ensuremath{\left(\delta^{il}\delta^{jk}-\delta^{ij}\delta^{kl}+\delta^{ik}\delta^{jl}\right)}}\label{A5}
\end{equation}
we obtain: 
\begin{equation}
\begin{split} & -2U_{7}\left(\sum G_{h}G_{e_{1}}G_{e_{2}}\right)^{2}M_{1}^{2}M_{2}^{2}\\
= & -\frac{U_{7}}{2}\left(\sum G_{h}G_{e_{1}}G_{e_{2}}\right)^{2}\left[\left(\mathbf{M}_{1}^{2}+\mathbf{M}_{2}^{2}\right)^{2}-\left(\mathbf{M}_{1}^{2}-\mathbf{M}_{2}^{2}\right)^{2}\right]
\end{split}
\label{A6}
\end{equation}

All the other terms can be computed in an analogous way.

\end{document}